\documentclass[twocolumn,epjc3]{svjour3}

\input{applfast-preamble}

\begin{document}

%-- overwrite "journal name" placeholder to add preprint #'s for arXiv submission
\renewcommand\makeheadbox{%
  \hfill
  CERN-TH-2022-125,
  IPPP/22/53,
  MPP-2022-80,
  ZU-TH 34/22
}
\title{NNLO interpolation grids for jet production at the LHC}

\author{%
  D.~Britzger\thanksref{mpi}\and
  A.~Gehrmann-De~Ridder\thanksref{eth,zurich}\and
  T.~Gehrmann\thanksref{zurich}\and
  E.W.N.~Glover\thanksref{durham}\and
  C.~Gwenlan\thanksref{oxford}\and
  A.~Huss\thanksref{cernth}\and
  J.~Pires\thanksref{lisboa1,lisboa2}\and
  K.~Rabbertz\thanksref{karls,cernex}\and
  D.~Savoiu\thanksref{hamburg}\and
  M.R.~Sutton\thanksref{susx}\and
  J.~Stark\thanksref{karls}
}

\raggedbottom

\institute{%
  Max-Planck-Institut für Physik, Föhringer Ring 6, D-80805 München, Germany \label{mpi}\and
  Institute for Theoretical Physics, ETH, Wolfgang-Pauli-Strasse 27, CH-8093 Zürich, Switzerland \label{eth}\and
  Physik-Institut, Universität Zürich, Winterthurerstrasse 190, CH-8057 Zürich, Switzerland \label{zurich}\and
  Institute for Particle Physics Phenomenology, Durham University, Durham, DH1 3LE, United Kingdom \label{durham}\and
  Department of Physics, The University of Oxford, Oxford, OX1 3PU, United Kingdom\label{oxford}\and
  Theoretical Physics Department, CERN, CH-1211 Geneva 23, Switzerland \label{cernth}\and
  LIP, Avenida Professor Gama Pinto 2, P-1649-003 Lisboa, Portugal \label{lisboa1}\and
  Faculdade de Ci\^encias, Universidade de Lisboa, 1749-016 Lisboa, Portugal \label{lisboa2}\and
  Institut für Experimentelle Teilchenphysik (ETP), KIT, Wolgang-Gaede-Str.\ 1, D-76131
  Karlsruhe, Germany\label{karls}\and
  Experimental Physics Department, CERN, CH-1211 Geneva 23, Switzerland \label{cernex}\and
  Institut für Experimentalphysik, Universität Hamburg, Luruper Chaussee 149, D-22761 Hamburg, Germany\label{hamburg}\and
  Department of Physics and Astronomy, The University of Sussex, Brighton, BN1 9RH, United Kingdom\label{susx}
}

\date{Date: \today}
\maketitle

% -----------------------------------------------------------------------
% \tableofcontents
% -----------------------------------------------------------------------

% -----------------------------------------------------------------------
%!TEX root = ../applfast.tex

% -----------------------------------------------------------------------
% abstract
% -----------------------------------------------------------------------

\begin{abstract}
  Fast interpolation-grid frameworks facilitate an efficient and flexible evaluation of higher-order predictions for any
  choice of parton distribution functions or value of the strong coupling \as.  They constitute an
  essential tool for the extraction of parton distribution functions and Standard Model parameters, as well as studies of the dependence 
  of cross sections on the renormalisation and factorisation scales.  The \APPLFAST project provides a generic interface between the parton-level
  Monte Carlo generator \NNLOJET and both the \APPLGRID and the \FASTNLO libraries for the grid interpolation.  The
  extension of the project to include hadron--hadron collider processes at next-to-next-to-leading order in perturbative QCD
  is presented, together  with an application for jet production at the LHC.
\end{abstract}

% -----------------------------------------------------------------------

% -----------------------------------------------------------------------
%!TEX root = ../applfast.tex

% -----------------------------------------------------------------------
\section{Introduction}
\label{intro}
% -----------------------------------------------------------------------

\sloppy
Theory predictions at next-to-next-to-leading order (NNLO) in perturbative QCD (pQCD) are the current new standard for an
increasingly large
range of important LHC processes~\cite{Heinrich:2020ybq}. This development is underscored by the completion of 
almost
all relevant $2\to2$ scattering processes at this order and first results for genuine $2\to3$
processes, see e.g.\ Ref.~\cite{Huss:2022ful} for a recent overview.
The \NNLOJET program~\cite{Gehrmann:2018szu} provides a single
framework for performing such calculations fully
differentially, and is under continuous development to provide state-of-the-art theory predictions for a plethora of processes.

Despite this remarkable progress, applications of these calculations, beyond simple
predictions, still remain the exception: computational efficiency is the primary bottleneck that restricts
the wide-spread use of NNLO calculations in applications such as the extraction of parton distribution functions (PDFs)~\cite{Amoroso:2022eow} or
Standard Model (SM) parameters such as \as~\cite{dEnterria:2022hzv} or a thorough assessment of theoretical uncertainties.
With a typical computing cost that exceeds $\mathcal{O}(10^5)$ CPU core hours, any application that relies on the repeated
calculation of the cross section with different input conditions---e.g.\ the variation of the value of the strong coupling \as,
the parametrisation of PDFs, or the study of the dependence on the renormalisation and factorisation scales---quickly becomes a
formidable challenge.  

The technique of fast interpolation grids~\cite{H1:2000bqr} addresses this bottleneck and has been implemented in the
\APPLGRID~\cite{Carli:2005ji,Carli:2010rw} and \FASTNLO~\cite{Kluge:2006xs,Britzger:2012bs} packages.  The extension of this
approach to NNLO predictions has been achieved for the case of DIS in Ref.~\cite{H1:2017bml,Britzger:2019kkb} and
diffractive DIS in Ref.~\cite{Britzger:2018zvv}.  First applications of the grid technique to hadron--hadron collision
processes at NNLO were discussed in
Refs.~\cite{BritzgerDIS14,Guzzi:2014wia,Czakon:2016dgf,Czakon:2017dip,Carrazza:2020gss}.  In this paper, the approach 
is extended to include the
NNLO processes in hadron--hadron collisions  
available from the \NNLOJET program.  The technique has been applied to the important and complex process of jet 
production at LHC energies, and is here 
described in detail.  The application to jet production cross sections can be seen as a
proof-of-principle for any further processes implemented in \NNLOJET at NNLO QCD.

The paper is structured as follows:  Section~\ref{sec:nnlo} provides a brief review of the grid technique, whilst
highlighting the main conceptual differences with respect to the DIS case and providing additional necessary details 
of the implementation;
the interpolation quality is discussed in Section~\ref{sec:incjet} and the inclusive jet production process is
used as an example.  Various sources of theoretical uncertainties are considered,  including those due to the scale, 
\as, and the PDF, followed by an assessment of the quality of the commonly employed \kfactor approach,
and an investigation of the total fiducial inclusive jet cross section at the LHC.  In
Section~\ref{sec:dijet}, a simultaneous \as and PDF fit is performed using the dijet process.  Grids are made
publicly available on the \PLOUGHSHARE website~\cite{web:ploughshare}.

% -----------------------------------------------------------------------

% -----------------------------------------------------------------------
%!TEX root = ../applfast.tex

% -----------------------------------------------------------------------
\section{NNLO predictions for hadron--hadron colliders and the \APPLFAST project}
\label{sec:nnlo}
% -----------------------------------------------------------------------

% -----------------------------------------------------------------------
\subsection{Differential predictions}
\label{sec:nnlo:diff}
% -----------------------------------------------------------------------

Cross section predictions for hadron--hadron collisions are described through QCD factorisation as a convolution of
the underlying hard scattering of partons and the
PDFs for each target hadron, 
\begin{align}
  \sigma =
  \int \rd x_1 \rd x_2
  & \;
    f_a(x_1,\muf)  f_b(x_2,\muf)
    \nonumber \\
  & \times
    \rd\hat{\sigma}_{ab}(x_1,x_2,\mur,\muf)
    \, ,
    \label{eq:sigma}
\end{align}
where an implicit summation over the incoming parton flavours $a$ and $b$ is assumed.  The hard-scattering cross section
can be obtained in pQCD as an expansion in the strong coupling
\begin{align*}
  &
    \rd\hat{\sigma}_{ab}(x_1,x_2,\mur,\muf) =
    \nonumber \\
  & \qquad
    \sum_{n=0}^{n_\mathrm{max}} \left(\frac{\as(\mur)}{2\pi}\right)^{p+n}
    \rd\hat{\sigma}^{(n)}_{ab}(x_1,x_2,\mur,\muf)
    \, ,
    % \label{eq:sigmahat_exp}
\end{align*}
where $p$ denotes the power in $\as$ of the leading-order (LO) process ($n=0$) and $n_\mathrm{max}$ the number of orders
beyond LO that are considered in the perturbative calculation.

Predictions beyond Born level ($n>0$) receive contributions that involve additional loop integrations and real-emission
corrections.  This gives rise to a set of parton-level ingredients of different particle multiplicities that are
individually divergent and only finite in their sum for sufficiently inclusive quantities.  Fully differential
predictions must retain the full kinematic information of the final state while at the same time ensuring the
cancellation of such infrared singularities.  The calculations within the \NNLOJET framework~\cite{Gehrmann:2018szu}
accomplish this task through the re-distribution of singularities using the antenna subtraction
formalism~\cite{GehrmannDeRidder:2005cm,Daleo:2006xa,Currie:2013vh}.  Arbitrary collinear- and infrared-safe observables
can be computed in this framework through a flexible parton-level Monte Carlo generator that samples the available
phase space $(x_{1;m},x_{2;m},\Phi_m)_{m=1,\ldots,M_n}$ with $M_n$ points and accumulates the associated weights
$w^{(n)}_{ab;m}$.  The cross section in Eq.~\eqref{eq:sigma} can then be computed via
\begin{align}
  \sigma \xrightarrow{\text{MC}}
  \sum_{n=0}^{n_\mathrm{max}} \sum_{m=1}^{M_n} &
  \;
  \left(\frac{\as(\mur{}_{;m})}{2\pi}\right)^{p+n}
  \nonumber\\&\times
  f_{a}(x_{1;m}, \muf{}_{;m}) \;
  f_{b}(x_{2;m}, \muf{}_{;m})
  \nonumber\\&\times
  w^{(n)}_{ab;m} \; \rd\hat{\sigma}^{(n)}_{ab;m} \, ,
  \label{eq:sigma_MC}
\end{align}
using the short-hand notation
\begin{align*}
  \mu_{X;m} &\equiv \mu_X(\Phi_m) \quad \text{for $X=\mathrm{R},\,\mathrm{F}$,}
  \\
  \rd\hat{\sigma}^{(n)}_{ab;m} &\equiv
  \rd\hat{\sigma}^{(n)}_{ab}(x_{1;m}, x_{2;m}, \mur{}_{;m},\muf{}_{;m}) \,.
\end{align*}
\NNLOJET further provides a decomposition of the logarithmic structure of the cross section
\begin{align}
  & \rd\hat{\sigma}^{(n)}_{ab} (\mur^2,\muf^2)
  = \sum_{\mathclap{\substack{\alpha, \beta \\ \alpha+\beta\leq n}}}
  \rd\hat{\sigma}^{(n | \alpha,\beta)}_{ab}
  \ln^{\alpha}\left(\frac{\mur^2}{\mu_0^2}\right)
  \ln^{\beta} \left(\frac{\muf^2}{\mu_0^2}\right)
  \label{eq:weight_decomp}
  \, ,
\end{align}
to facilitate a flexible reconstruction of the scale dependence. 
Here, $\mu_0$ denotes an arbitrary hard reference
scale used for this decomposition.

% -----------------------------------------------------------------------
\subsection{The grid technique for hadron--hadron collisions}
\label{sec:nnlo:grid}
% -----------------------------------------------------------------------

The grid technique for hadron--hadron collisions at NNLO is a non-trivial extension of that for DIS~\cite{Britzger:2019kkb} taking
into account an additional parton distribution for the second target hadron and the corresponding momentum fraction,
$x_2$.  Adopting the same notation as in Ref.~\cite{Britzger:2019kkb}, the equivalent of Eq.~(11) from there for the hadron--hadron
case becomes
\begin{align}
  &
  \as(\mu) \; f_a(x_1,\mu)  f_b(x_2,\mu) \simeq
  \nonumber \\
  & \qquad
  \sum_{i,j,k} \as^{[k]} \;
  f^{[i,k]}_{a} f^{[j,k]}_{b} \;
  E^y_i(x_1)  E^y_j(x_2) \; E^\tau_k(\mu)
  \, ,
  \label{eq:kernels}
\end{align}
where $\mur=\muf\equiv\mu$ has been set for simplicity.  The summation over $i$, $j$, and $k$ represents the summation
over the nodes of the grid structure for $x_1$, $x_2$, and $\mu$, respectively, where one dimension in the grid is needed 
for each interpolated parameter.
The superscripts on the
interpolation kernels $E_i^y(x)$ denote variable transformations $x \longmapsto y(x)$ that are introduced to allow for a 
more optimal span of the phase space of the transformed variable and to improve
the interpolation quality with respect to using equidistant grid nodes.  Some common choices for the transformations are given
explicitly in Ref.~\cite{Britzger:2019kkb}.

The na\"{i}ve application of the sum over the parton flavours in Eq.~\eqref{eq:kernels} results in up to
121 ($11\times11$) different parton--parton luminosity contributions, or 169 if the top quark is also included, which
makes the representation as a numerical grid excessively large and potentially prohibitive for practical applications.
It is therefore expedient to instead make use of symmetries within the structure of the hard subprocesses to form linear
combinations of the individual parton--parton luminosities to arrange for a smaller set of unique luminosities.  This
allows the summation over the full set of parton flavour combinations ($a$ and $b$) to be replaced by a single summation
over a significantly smaller set of contributions, $F_\lambda(x_1,x_2,\mu)$, such that
\begin{align}
  &
  \sum_\lambda F_\lambda(x_1,x_2,\mu) \;
  h_\lambda(x_1,x_2, \mu) \equiv
  \nonumber \\
  & \qquad
  \sum_{a,b} f_a(x_1,\mu)  f_b(x_2,\mu) \;
  h_{ab}(x_1,x_2, \mu),
\end{align}
where the summations over parton luminosities have been included explicitly on this occasion.  The weights (denoted as
$h$) for any specific contribution $\lambda$ are identical for each of the individual terms $(a,b)$ in the summation for
the corresponding index $\lambda$.
As an example, the decomposition within \NNLOJET for jet production in hadron--hadron collisions~\cite{Currie:2018xkj,Currie:2017eqf,Gehrmann-DeRidder:2019ibf,Chen:2022tpk} is
\begingroup \allowdisplaybreaks
\begin{equation}
\begin{aligned}
  % [0, 0]
  F_{1} &=
  f_{g}(x_1) \,
  f_{g}(x_2)
  ,
  \\
  % -----
  % [30, 0]
  F_{2} &=
  \sum_{i=1}^{6}
  f_{q_i}(x_1) \,
  f_{g}(x_2)
  , &
  % [-30, 0]
  F_{\bar{2}} &=
  \sum_{i=1}^{6}
  f_{\bar{q}_i}(x_1) \,
  f_{g}(x_2)
  \\*
  % [0, 30]
  F_{3} &=
  \sum_{i=1}^{6}
  f_{g}(x_1) \,
  f_{q_i}(x_2)
  , &
  % [0, -30]
  F_{\bar{3}} &=
  \sum_{i=1}^{6}
  f_{g}(x_1) \,
  f_{\bar{q}_i}(x_2)
  \\
  % -----
  % [30, -30]
  F_{4} &=
  \sum_{i=1}^{6}
  f_{q_i}(x_1) \,
  f_{\bar{q}_i}(x_2)
  , &
  % [-30, 30]
  F_{\bar{4}} &=
  \sum_{i=1}^{6}
  f_{\bar{q}_i}(x_1) \,
  f_{q_i}(x_2)
  \\*
  % [30, 30]
  F_{5} &=
  \sum_{i=1}^{6}
  f_{q_i}(x_1) \,
  f_{q_i}(x_2)
  , &
  % [-30, -30]
  F_{\bar{5}} &=
  \sum_{i=1}^{6}
  f_{\bar{q}_i}(x_1) \,
  f_{\bar{q}_i}(x_2)
  \\
  % -----
  % [31, -31]
  F_{6} &=
  \sum_{i,j=1}^{6}
  f_{q_i}(x_1) \,
  f_{\bar{q}_j}(x_2)
  , &
  % [-31, 31]
  F_{\bar{6}} &=
  \sum_{i,j=1}^{6}
  f_{\bar{q}_i}(x_1) \,
  f_{q_j}(x_2)
  \\*
  % [31, 31]
  F_{7} &=
  \sum_{i,j=1}^{6}
  f_{q_i}(x_1) \,
  f_{q_j}(x_2)
  , &
  % [-31, -31]
  F_{\bar{7}} &=
  \sum_{i,j=1}^{6}
  f_{\bar{q}_i}(x_1) \,
  f_{\bar{q}_j}(x_2)
  ,
\end{aligned}
\end{equation}
\endgroup
effectively reducing the number of separate contributions that must be stored in the grid from up to 121 down
to 13.  This reduction of the parton luminosities is automatically performed in \APPLFAST for any hadron--hadron process
based on the process-dependent implementation in \NNLOJET. For jet production this number could in principle be
further reduced down to 7 independent combinations~\cite{Carli:2010rw}.

Using this reduced number of parton luminosities, the interpolated cross section prediction can be written as
\begin{align}
  \sigma &\simeq
  \sum_{n} \sum_{i,j,k} \biggl(\frac{\as^{[k]}}{2\pi}\biggr)^{p+n}
  F^{[i,j,k]}_{\lambda} \;
  \hat{\sigma}^{(n)}_{\lambda[i,j,k]}
  \, ,
  \label{eq:sigma_grid}
\end{align}
where the summation over $\lambda$ is implied.
The corresponding grid is obtained by accumulating the weights according to
\begin{align}
  &
  \hat{\sigma}^{(n)}_{\lambda[i,j,k]} \xrightarrow{\text{MC}}
  \nonumber\\
  & \qquad
  \sum_{m=1}^{M_n}
  E^y_i(x_{1;m})
  E^y_j(x_{2;m})
  E^\tau_k(\mu_{m})
  \; w^{(n)}_{\lambda;m}
  \; \rd\hat{\sigma}^{(n)}_{\lambda;m}
  \, ,
  \label{eq:grid_gen}
\end{align}
where now the terms $w^{(n)}_{\lambda;m}$ correspond to those weights $w^{(n)}_{ab;m}$ associated with the individual
terms for $\lambda$.

% -----------------------------------------------------------------------
\subsection{Renormalisation and factorisation scale dependence}
\label{sec:nnlo:scale}
% -----------------------------------------------------------------------

A flexible variation of the renormalisation and factorisation scales in NNLO pQCD predictions is important for many
phenomenological applications. The interpolation grids developed here allow the variation of the scales by
arbitrary factors, and a selection of different scale choices, without a recalculation of the hard coefficients.  With the hard
coefficients $\hat{\sigma}^{(n)}_{\lambda[i,j,k]}$ determined separately order by order in $\as$, the dependence on the
renormalisation and factorisation scales, $\mur$ and $\muf$, can be restored using the RGE running of $\as$ and the
DGLAP evolution for the PDFs.  Introducing a generic functional form depending on the scale choice $\mu$ during the grid
generation  in Eq.~\eqref{eq:grid_gen},
\begin{align}
  \mu_X &=
  \mu_X(\mu) \quad \text{for $X=\mathrm{R},\,\mathrm{F}$}
  \,,
\end{align}
and using the short-hand notation from Ref.~\cite{Britzger:2019kkb},
\begin{align*}
  L^{[k]}_{\mathrm{X}} &\equiv
  \ln\left(\frac{\mu_X^2(\mu^{[k]})}{\mu^{2[k]}}\right)
  \quad \text{for $X=\mathrm{R},\,\mathrm{F}$},
  \nonumber\\
  \as^{[k_{\to\mathrm{R}}]} &\equiv
  \as(\mur(\mu^{[k]})) \, ,
  \quad \text{and} \\
  F^{[i,j,k_{\to\mathrm{F}}]}_{\lambda}  &\equiv
  F_\lambda(x_1^{[i]}, x_2^{[j]},\muf(\mu^{[k]})) \, ,
\end{align*}
the full scale dependence up to NNLO is then given by~\cite{Currie:2018xkj}
\begingroup \allowdisplaybreaks
\begin{align}
  &\sigma^\text{NNLO}(\mur,\muf) =
  \sum_{i,j,k} \biggl(\frac{\as^{[k_{\to\mathrm{R}}]}}{2\pi}\biggr)^{p}
  F^{[i,j,k_{\to\mathrm{F}}]}_\lambda \; \hat{\sigma}^{(0)}_{\lambda[i,j,k]}
  \nonumber\\&\quad
  +\sum_{i,j,k} \biggl(\frac{\as^{[k_{\to\mathrm{R}}]}}{2\pi}\biggr)^{p+1}
  \biggl\{
  F^{[i,j,k_{\to\mathrm{F}}]}_\lambda \; \hat{\sigma}^{(1)}_{\lambda[i,j,k]}
  \nonumber\\*&\qquad
  + \Bigl[
  p \beta_0 F^{[i,j,k_{\to\mathrm{F}}]}_\lambda L^{[k]}_{\mathrm{R}}
  \nonumber\\*&\qquad\quad
  -\left(
  F_{\lambda;{f_a\to P_0\otimes f_a}}^{[i,j,k_{\to\mathrm{F}}]}
  + F_{\lambda;{f_b\to P_0\otimes f_b}}^{[i,j,k_{\to\mathrm{F}}]}
  \right) L^{[k]}_{\mathrm{F}}
  \Bigr] \; \hat{\sigma}^{(0)}_{\lambda[i,j,k]}
  \biggr\}
  \nonumber\\&\quad
  +\sum_{i,j,k} \biggl(\frac{\as^{[k_{\to\mathrm{R}}]}}{2\pi}\biggr)^{p+2}
  \biggl\{
  F_\lambda^{[i,j,k_{\to\mathrm{F}}]} \; \hat{\sigma}^{(2)}_{\lambda[i,j,k]}
  \nonumber\\&\qquad
  + \Bigl[
  (p+1) \beta_0 F^{[i,j,k_{\to\mathrm{F}}]}_\lambda L^{[k]}_{\mathrm{R}}
  \nonumber\\*&\qquad\quad
  \;-\left(
  F_{\lambda;{f_a\to P_0\otimes f_a}}^{[i,j,k_{\to\mathrm{F}}]}
  +F_{\lambda;{f_b\to P_0\otimes f_b}}^{[i,j,k_{\to\mathrm{F}}]}
  \right) L^{[l]}_{\mathrm{F}}
  \Bigr] \; \hat{\sigma}^{(1)}_{\lambda[i,j,k]}
  \nonumber\\&\qquad
  + \Bigl[
  \Bigl(
  p \beta_1 + \tfrac{1}{2}p(p+1)\beta_0^2 L^{[k]}_{\mathrm{R}}
  \Bigr) \; F^{[i,j,k_{\to\mathrm{F}}]}_\lambda L^{[j]}_{\mathrm{R}}
  \nonumber\\*&\qquad\quad
  \;- \left(
  F_{\lambda;{f_a\to P_1\otimes f_a}}^{[i,j,k_{\to\mathrm{F}}]}
  +F_{\lambda;{f_b\to P_1\otimes f_b}}^{[i,j,k_{\to\mathrm{F}}]}
  \right) L^{[k]}_{\mathrm{F}}
  \nonumber\\*&\qquad\quad
  \;+\tfrac{1}{2}
  \left(
  F_{\lambda;{f_a\to P_0\otimes P_0\otimes f_a}}^{[i,j,k_{\to\mathrm{F}}]}
  + F_{\lambda;{f_b\to P_0\otimes P_0\otimes f_b}}^{[i,j,k_{\to\mathrm{F}}]}
  \right)  L^{2[k]}_{\mathrm{F}}
  \nonumber\\*&\qquad\quad
  \,+ \Bigl(
  \tfrac{1}{2} \beta_0 L^{[j]}_{\mathrm{F}}  -(p+1) \beta_0 L^{[k]}_{\mathrm{R}}
  \Bigr)
  \nonumber\\*&\qquad\quad\quad
  \;\times
  \left(
  F_{\lambda;{f_a\to P_0\otimes f_a}}^{[i,j,k_{\to\mathrm{F}}]}
  +F_{\lambda;{f_b\to P_0\otimes f_b}}^{[i,j,k_{\to\mathrm{F}}]}
  \right)
  L^{[k]}_{\mathrm{F}}
  \nonumber\\&\qquad\quad
  \;+ F_{\lambda;{f_a\to P_0\otimes f_a};{f_b\to P_0\otimes f_b}}^{[i,j,k_{\to\mathrm{F}}]} L^{2[k]}_{\mathrm{F}}
  \Bigr] \; \hat{\sigma}^{(0)}_{\lambda[i,j,k]}
  \biggr\}
  \, .
  \label{eq:scale_ana}
\end{align}
\endgroup
Here, the notation $F_{\lambda;{f_a\to X_a}}$ represents the term $F_\lambda$ but with $f_a$
replaced by $X_a$.
In \APPLGRID, the calculation of the scale-dependent terms is performed only if and when required, with the 
convolutions involving the splitting functions $P_{n}$ evaluated using \HOPPET~\cite{Salam:2008qg}.

As an alternative to the analytical reconstruction of the scale variation in Eq.~\eqref{eq:scale_ana}, additional 
individual grids for each scale-independent coefficient can be generated, which then are multiplied with the scale-dependent logarithms.
This corresponds to the default strategy in the \FASTNLO library where the full scale dependence is reconstructed
using
\begingroup \allowdisplaybreaks
\begin{align}
  &\sigma^\text{NNLO}(\mur,\muf) =
    \sum_{i,j,k} \biggl(\frac{\as^{[k_{\to\mathrm{R}}]}}{2\pi}\biggr)^{p}
    F^{[i,j,k_{\to\mathrm{F}}]}_{\lambda} \; \hat{\sigma}^{(0|0,0)}_{\lambda[i,j,k]}
    \nonumber\\&\quad
  +\sum_{i,j,k} \biggl(\frac{\as^{[k_{\to\mathrm{R}}]}}{2\pi}\biggr)^{p+1}
  F^{[i,j,k_{\to\mathrm{F}}]}_{\lambda} \;
  \nonumber\\*&\quad \quad\times
  \biggl\{
  \hat{\sigma}^{(1|0,0)}_{\lambda[i,j,k]}
  + L^{[k]}_{\mathrm{R}} \; \hat{\sigma}^{(1|1,0)}_{\lambda[i,j,k]}
  + L^{[k]}_{\mathrm{F}} \; \hat{\sigma}^{(1|0,1)}_{\lambda[i,j,k]}
  \biggr\}
  \nonumber\\&\quad
  +\sum_{i,j,k} \biggl(\frac{\as^{[k_{\to\mathrm{R}}]}}{2\pi}\biggr)^{p+2}
  F^{[i,j,k_{\to\mathrm{F}}]}_{\lambda} \;
  \nonumber\\*&\quad \quad\times
  \biggl\{
  \hat{\sigma}^{(2|0,0)}_{\lambda[i,j,k]}
  + L^{[k]}_{\mathrm{R}} \; \hat{\sigma}^{(2|1,0)}_{\lambda[i,j,k]}
  + L^{[k]}_{\mathrm{F}} \; \hat{\sigma}^{(2|0,1)}_{\lambda[i,j,k]}
  \nonumber\\*&\qquad\quad
  + L^{2[k]}_{\mathrm{R}} \; \hat{\sigma}^{(2|2,0)}_{\lambda[i,j,k]}
  + L^{2[k]}_{\mathrm{F}} \; \hat{\sigma}^{(2|0,2)}_{\lambda[i,j,k]}
  \nonumber\\*&\qquad\quad
  + L^{[k]}_{\mathrm{R}} \; L^{[k]}_{\mathrm{F}} \; \hat{\sigma}^{(2|1,1)}_{\lambda[i,j,k]}
  \biggr\}
  \, ,
  \label{eq:scale_log}
\end{align}
\endgroup
where the grids are produced in analogy to Eq.~\eqref{eq:grid_gen} but using the decomposition of
Eq.~\eqref{eq:weight_decomp}
\begin{align*}
  &
    \hat{\sigma}^{(n|\alpha,\beta)}_{\lambda[i,j]} \xrightarrow{\text{MC}}
    \nonumber\\
  & \qquad
    \sum_{m=1}^{M_n}
    E^y_i(x_{1;m})
    E^y_j(x_{2;m})
    E^\tau_k(\mu_{m})
    \; w^{(n)}_{\lambda;m}
    \; \rd\hat{\sigma}^{(n|\alpha,\beta)}_{\lambda;m}
    \, .
\end{align*}

% -----------------------------------------------------------------------

% -----------------------------------------------------------------------
%!TEX root = ../applfast.tex

% -----------------------------------------------------------------------
\section{Applications of interpolation grids for inclusive jet production}
\label{sec:incjet}
% -----------------------------------------------------------------------

This section presents an example of the analyses that can be performed using interpolation grids for
inclusive jet production cross sections at the LHC.  
Measurements at 7\,\TeV by ATLAS~\cite{ATLAS:2014riz} are used; the quality of the jet
interpolation grids is demonstrated in Section~\ref{sec:incjet:closure}, and in Section~\ref{sec:incjet:unc} the grids are used to perform 
a detailed comparison of scale, \as, and PDF uncertainties. 
Section~\ref{sec:incjet:k-fac} studies the robustness of the NNLO \kfactor approach that is commonly
used in PDF fits as a proxy for the exact NNLO prediction.
The grids are subsequently employed for a detailed investigation of the total
inclusive jet cross section, where NNLO predictions are compared to
measurements from both ATLAS and CMS.

% -----------------------------------------------------------------------
\subsection{Interpolation grids for inclusive jet production}
\label{sec:incjet:grids}
% -----------------------------------------------------------------------

Inclusive jet production cross sections for the anti-\kt~\cite{Cacciari:2008gp} 
jet algorithm have been measured in proton--proton collisions by the ATLAS and CMS 
collaborations at different centre-of-mass energies and for different values of the
jet-radius ($R$) parameter.  Interpolation grids at NNLO have been generated 
for a large selection of these measurements and 
a summary of the NNLO predictions, together with their respective kinematic ranges, 
is provided in Table~\ref{tab:incjet:datasets}.%
\footnote{
  Note that the predictions here provided are based on the leading-colour and leading $N_f$ approximation and do not include the most recent sub-leading colour contributions presented in Ref.~\cite{Chen:2022tpk}.
}
%
% ----- Inclusive jet datasets
\begin{table*}
  \caption{An overview of inclusive jet \pt datasets with \APPLFAST interpolation grids for 
    proton--proton collisions at the LHC\@.
    For each dataset the centre-of-mass energy $\sqrt{s}$, the integrated luminosity $\mathcal{L}$,
    the number of data points, and the jet algorithm are listed.
    Jets are required to be within a given range of rapidity $y$ in the
    laboratory frame.  Available choices for the central scales for \murf for the interpolation grids are listed.%
  }
  \label{tab:incjet:datasets}
  \scriptsize\renewcommand{\arraystretch}{1.3}
  \begin{center}
    % \begin{tabular}{l@{\hskip4pt}c@{\hskip4pt}c@{\hskip2pt}cccccccc@{\hskip0pt}}
    \begin{tabular}{l@{\hskip4pt}c@{\hskip4pt}c@{\hskip2pt}ccccc@{\hskip0pt}}
      \toprule
      \textbf{Data} & \textbf{\boldmath$\sqrt{s}$} & \textbf{\boldmath{$\mathcal{L}$}}
      & \textbf{no.\ of} & \textbf{anti-\boldmath{\kt}} & \textbf{kinematic range} & \textbf{fiducial cuts}
      & \textbf{\boldmath\murf-choice}
      % & \textbf{num.} & \textbf{stat.} & \textbf{CPU time}
      \\
      & \textbf{[TeV]} & \textbf{\boldmath{$[{\rm fb}^{-1}]$}}
      & \textbf{points} & \textbf{\boldmath{$R$}} & \textbf{[GeV]} & &
      % & \textbf{acc. [\%]} & \textbf{prec. [\%]} & \textbf{[kh]}
      \\
      \midrule % arXiv:1512.06212, Inspire:
      CMS~\cite{CMS:2015jdl} & %
      2.76 & 0.00543 & 81 &
      $0.7$ & $\ptjet\in[74,592]$ & $|y|<3.0$ &
      \ptjet,\htp
      % & $<0.1$ & $<1 \nearrow 3$ & 164
      \\
      \midrule % arXiv:1410.8857, Inspire:1325553
      ATLAS~\cite{ATLAS:2014riz} & %
      7.0 & 4.5 & $140$ &
      $0.6$ & $\ptjet\in[100,1992]$ & $|y|<3.0$ &
      \ptjet,\htp
      % &  &  &
      \\
      \midrule % arXiv:1212.6660, Inspire:
      CMS~\cite{CMS:2012ftr} & % fnl2332d
      7.0 & 5.0 & 133 &
      $0.7$ & $\ptjet\in[114,2116]$ & $|y|<3.0$ &
      \ptjet,\htp
      % & $<0.1$ & $<1 \nearrow 2$ & 156
      \\
      \midrule % arXiv:1706.03192, Inspire:1604271
      ATLAS~\cite{ATLAS:2017kux} & %
      8.0 & 20.3 & $171$ &
      $0.6$ & $\ptjet\in[70,2500]$ & $|y|<3.0$ &
      \ptjet,\htp
      % & & &
      \\
      \midrule % arXiv:1609.05331, Inspire:
      \multirow{2}{*}{CMS~\cite{CMS:2016lna}} & % fnl3332e
      \multirow{2}{*}{8.0} & 5.6 & \multirow{2}{*}{248} &
      \multirow{2}{*}{$0.7$} & $\ptjet\in[21,74]$ & \multirow{2}{*}{$|y|<4.7$} &
      \multirow{2}{*}{\ptjet,\htp}
      % & \multirow{2}{*}{$<0.1 \nearrow 0.5$} & \multirow{2}{*}{$<1 \nearrow 4$} & \multirow{2}{*}{250}
      \\
      && 19.7 &&& $\ptjet\in[74,2500]$ & &
      % & & &
      \\
      \midrule % arXiv:1711.02692, Inspire:1634970
      ATLAS~\cite{ATLAS:2017ble} & %
      13.0 & 3.2 & 177 &
      $0.4$ & $\ptjet\in[100,3937]$ & $|y|<3.0$ &
      \ptjet,\htp
      % & & &
      \\
      \midrule % arXiv:2111.10431, Inspire:
      \multirow{2}{*}{CMS~\cite{CMS:2021yzl}} & % fnl53(6,3)]2h
      \multirow{2}{*}{13.0} & 36.3 & \multirow{2}{*}{$2\times 78$} &
      $0.4$ & \multirow{2}{*}{$\ptjet\in[97,3103]$} & \multirow{2}{*}{$|y|<2.0$} &
      \multirow{2}{*}{\ptjet,\htp}
      % & \multirow{2}{*}{$<0.1 \nearrow 0.2$} & \multirow{2}{*}{$<1 \nearrow 4$} & 201
      \\
      && 33.5 && $0.7$ &&&
      % &&& 213
      \\
      \bottomrule
    \end{tabular}
  \end{center}
\end{table*}
For each calculation, a dedicated optimisation was employed using
kinematic reweighting factors and adaptation of the phase space integration.  
In order to achieve a sufficient numerical accuracy, the typical statistical
precision of the Monte Carlo integration is smaller than 1\% in most bins, with
exceptions only for some bins near the edges of the selected phase space.  
On modern batch computing systems, these calculations typically require between  
$3\cdot10^5$ and $6\cdot10^5$ CPU hours.  Together with the generation of the interpolation grids, the usual 
cross sections from \NNLOJET are still calculated and can be used as a reference to allow numerical
closure tests.  The calculations are performed using either the NNPDF3.1~\cite{Ball:2017nwa}
or the CT14~\cite{Dulat:2015mca} PDF sets at NNLO. 
Central scale choices of $\ptjet$ and $\htp$ are available, encoded in a single interpolation grid in the case of 
\FASTNLO, or available in separate grids in the case of \APPLGRID.

% -----------------------------------------------------------------------
\subsection{Closure test}
\label{sec:incjet:closure}
% -----------------------------------------------------------------------

As an important first step in the use of the interpolation grids, the degree of consistency between the NNLO predictions obtained
from the grids and the raw \NNLOJET prediction is studied.  In this case, the steps in the \APPLFAST procedure to generate
grids remain identical with the DIS case, described in Section~4 of Ref.~\cite{Britzger:2019kkb}.  During the 
execution of the calculation, in addition to
the interpolation grids themselves, the usual reference predictions from \NNLOJET are produced, which correspond to 
a computation solely based on \NNLOJET without grid generation.  For the consistency comparison of the cross section 
from the grid convolution with the reference cross section, the fast convolution uses the same scale and PDF choice 
as for the original calculation. This helps to ensure, for instance, that the density of interpolation nodes is
sufficient, such that interpolation errors are negligible in comparison to the statistical uncertainty of the NNLO
prediction.

\begin{figure}
  \centering
  \includegraphics[scale=1.2]{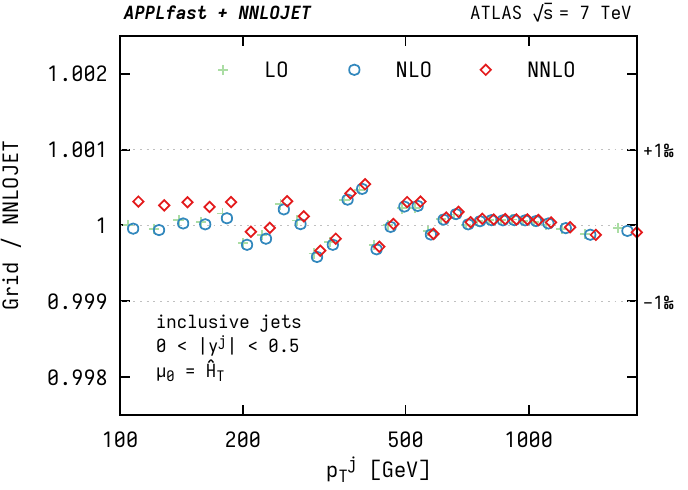}
  \\\medskip
  \includegraphics[scale=1.2]{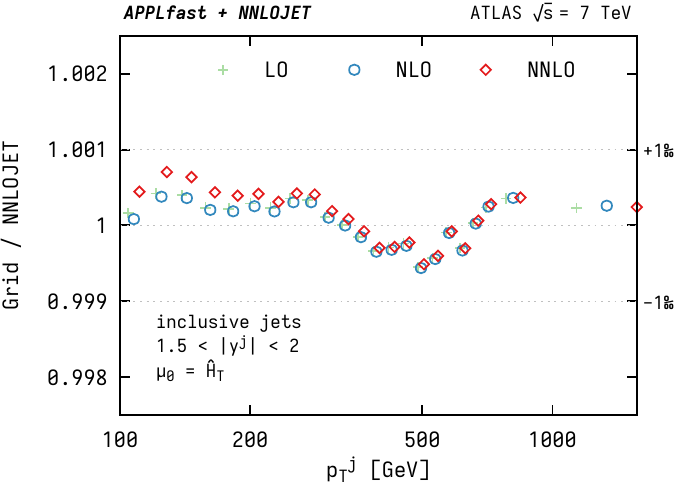}
  \caption{\label{fig:closure}%
    Closure tests for the ATLAS 7\,TeV inclusive jet cross section grid as a function of \ptjet for two representative ranges
    in rapidity. The horizontal dotted lines indicate the targeted closure of one permille.}
\end{figure}

Closure tests for the inclusive jet \pt distribution are shown in Fig.~\ref{fig:closure} for two 
representative ranges in jet rapidity. 
It can be seen that the interpolation tables are able to reproduce the reference prediction at
permille level, which is well below the experimental uncertainty and  the residual Monte Carlo statistical uncertainty of the
calculation itself.  Slight systematic trends are observed, where the interpolation quality degrades
somewhat towards more forward rapidity.  This is to be expected as the forward region samples a wider range of the momentum
fractions, $x_i$, thus introducing larger interpolation errors when the number of nodes is the same for all bins.  In
principle, the density of interpolation nodes can be adjusted in a phase-space dependent manner to mitigate this
degradation, but this would result in larger file sizes and slower grid convolution. 
Closure tests were performed for all grids that are made available together with this publication,   
all of which show a similar level of interpolation quality that is  
typically below the permille level and reaches a maximum of 0.2\% in exceptional phase-space regions.

% -----------------------------------------------------------------------
\subsection{Scale, \as, and PDF variations and their uncertainties}
\label{sec:incjet:unc}
% -----------------------------------------------------------------------

\begin{figure*}
  \includegraphics[scale=1.2]{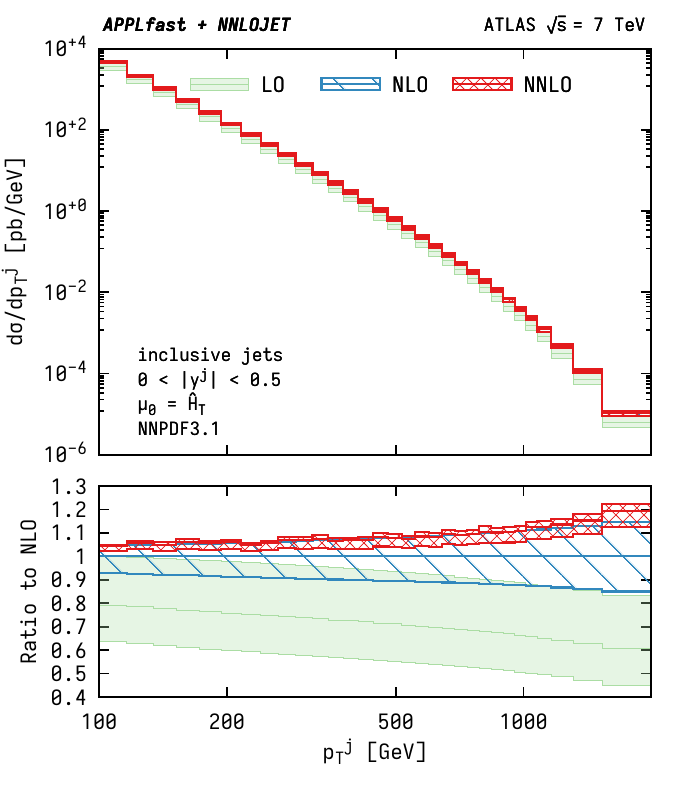}\hfill
  \includegraphics[scale=1.2]{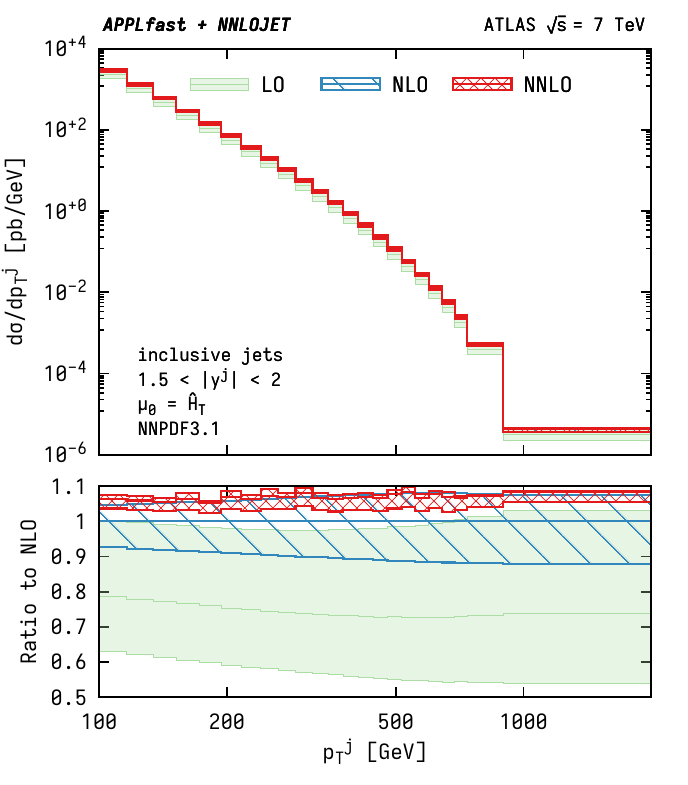}
  \caption{\label{fig:scl_unc}%
    Predictions at LO, NLO and NNLO for the 
    ATLAS 7\,TeV inclusive jet cross section as a function of \ptjet for two representative ranges in rapidity. 
    The shaded bands show the scale uncertainties when using the recommended scale
    choice of $\mu=\htp$ for inclusive jets.
    The lower panel displays the ratio with respect to the NLO prediction.}
\end{figure*}

The predictions up to NNLO using the newly generated grids are shown in Fig.~\ref{fig:scl_unc} 
for two representative rapidity ranges chosen for illustration. The bands correspond to the envelope from independently
varying \mur and \muf up and down by factors of two with the constraint $\tfrac{1}{2}\leq\mur/\muf\leq2$.
A small resulting spread of the calculations is observed, with the successive orders displaying a satisfactory 
overlap within their respective uncertainty estimates and, importantly, a dramatic reduction in the width of the 
scale variation bands is observed to result from the inclusion of  higher-order corrections.

\begin{figure*}
  \includegraphics[scale=1.35,trim={0 0 50 0},clip]{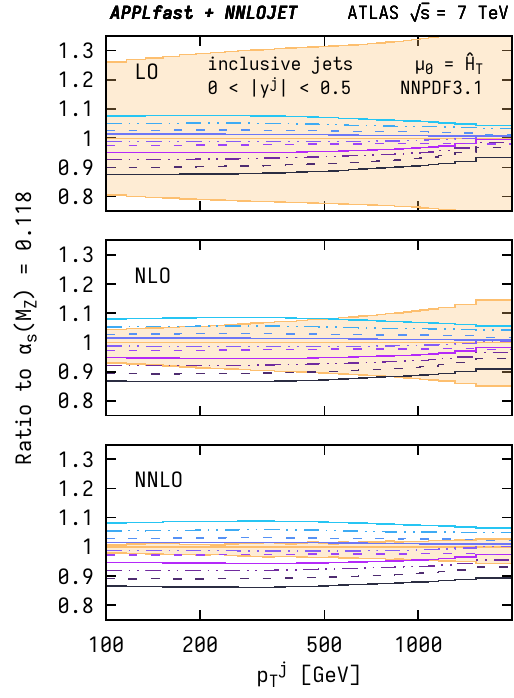}\hfill
  \includegraphics[scale=1.35]{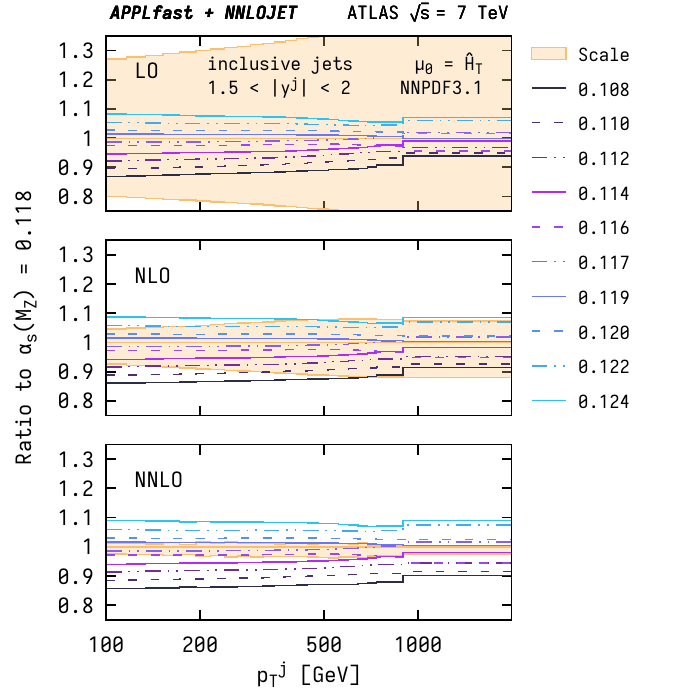}
  \caption{\label{fig:alps_unc}%
    LO, NLO, and NNLO predictions for the ATLAS 7\,TeV inclusive jet cross section as a function of \ptjet for two representative ranges in rapidity 
    and for different values of \asmz. 
    The predictions are obtained with \as-dependent variants of the NNPDF3.1 PDF set. The shaded area indicates the
    scale uncertainty at the given perturbative order.
  }
\end{figure*}

The sensitivity of the jet spectrum to the strong coupling, \as, is presented in Fig.~\ref{fig:alps_unc} for the different
perturbative orders.  To this end, the NNPDF3.1 PDF sets with different values for the strong coupling are used,
covering the range between $0.108 \leq \asmz \leq 0.124$.
It is observed that the scale variation at LO is substantially larger than that from varying \as, 
while, at NLO, the spread in predictions from each of these two sources are similar. 
In contrast, the reduced scale-variation uncertainty at NNLO allows for the resolution of the variation 
of \as at the level of a few percent for the first time,
illustrating the need for at least NNLO predictions for a robust extraction of \as at this level of precision.

\begin{figure*}
  \includegraphics[scale=1.35,trim={0 0 50 0},clip]{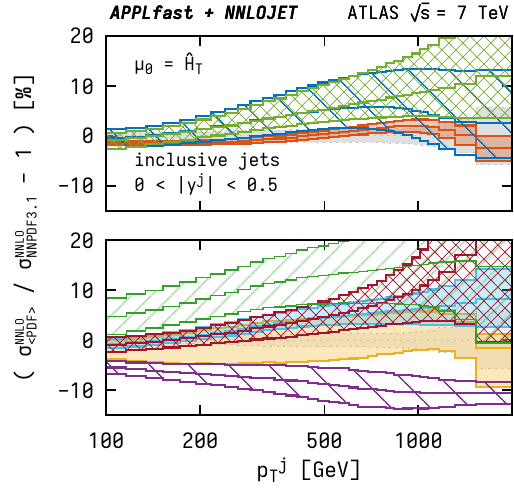}\hfill
  \includegraphics[scale=1.35]{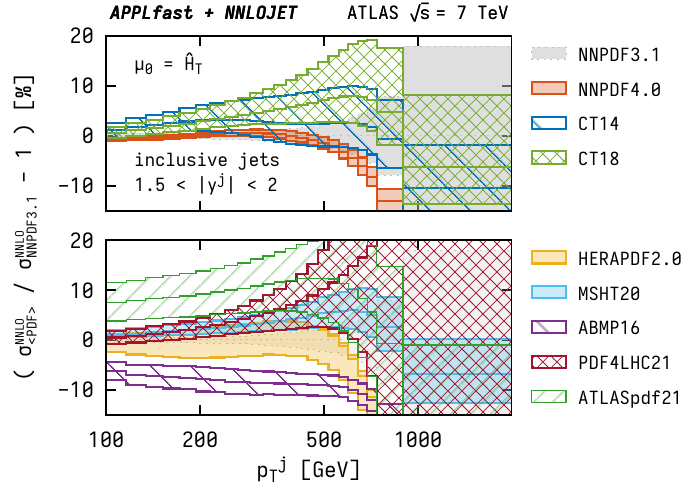}
  \caption{\label{fig:pdf_unc}%
    NNLO predictions for the ATLAS 7\,TeV inclusive jet cross section as a function of \ptjet 
    for two representative ranges in rapidity  
    and for various different PDF sets. The shaded areas indicate the respective PDF uncertainties.}
\end{figure*}

The fast convolution that is made possible by using the grids allows the provision of NNLO predictions with different PDF sets,
together with their complete respective uncertainties.  Figure~\ref{fig:pdf_unc} contrasts the NNLO prediction 
using the NNPDF3.1~\cite{Ball:2017nwa}, NNPDF4.0~\cite{NNPDF:2021njg}, CT14~\cite{Dulat:2015mca},
CT18~\cite{Hou:2019efy}, HERAPDF2.0~\cite{H1:2015ubc}, MSHT20~\cite{Bailey:2020ooq},
ABMP16~\cite{Alekhin:2017kpj}, PDF4LHC21~\cite{Ball:2022hsh}, and ATLASpdf21~\cite{ATLAS:2021vod} PDF sets by showing the
relative difference with respect to the prediction using the NNPDF3.1 set.
With the exception of the ABMP16,
HERAPDF2.0, and ATLASpdf21 sets, the predictions for the different PDF sets are mutually compatible within their respective uncertainty
estimates.  The PDF uncertainties are typically at the level of a few percent in the low-\pt regime and increase towards
larger \pt (or rapidity $y$) to $\mathcal{O}(10\%)$ uncertainties in the \TeV range.

% -----------------------------------------------------------------------
\subsection{Robustness of NNLO \texorpdfstring{\kfactors}{K-factors}}
\label{sec:incjet:k-fac}
% -----------------------------------------------------------------------

Owing to the large computational expenditure that NNLO calculations for jet production entail, their direct 
use in PDF fits has previously been unfeasible.  Instead, a common approach proceeds by complementing predictions 
from NLO interpolation grids with an NNLO \kfactor~\cite{AbdulKhalek:2020jut}.
The NNLO \kfactor is a proxy for the full NNLO prediction and defined as
\begin{align}
  K^\NNLO (\mu) &\equiv
  \frac{\rd\sigma^\NNLO(\mu)/\rd\pt}{\rd\sigma^\NLO(\mu)/\rd\pt}
  \, ,
  \label{eq:Kfac}
\end{align}
where the dependence on $(\mur,\muf)$ is abbreviated with a single scale $\mu$ for simplicity.  To this end, both the
numerator and denominator are evaluated with the same PDF set.  However, the NNLO \kfactor can be applied in 
two different ways, namely
\begin{subequations}
\begin{align}
  \sigma_\text{approx.\,1}^{\NNLO} (\mu) &=
  \sigma^{\NLO} (\mu) \times K^\NNLO (\muref)
  \quad\text{or}
  \label{eq:NNLO_approx1}
  \\
  \sigma_\text{approx.\,2}^{\NNLO} (\mu) &=
  \sigma^{\NLO} (\mu) \times K^\NNLO (\mu)
  \, ,
  \label{eq:NNLO_approx2}
\end{align}
\label{eq:NNLO_approx}
\end{subequations}
the consequences of which will be discussed in the remainder of this section.

\begin{figure*}
  \includegraphics[scale=1.35,trim={0 0 50 0},clip]{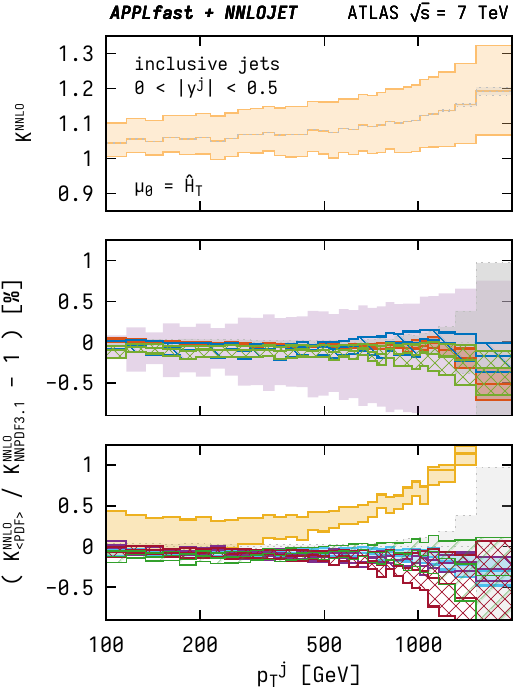}\hfill
  \includegraphics[scale=1.35]{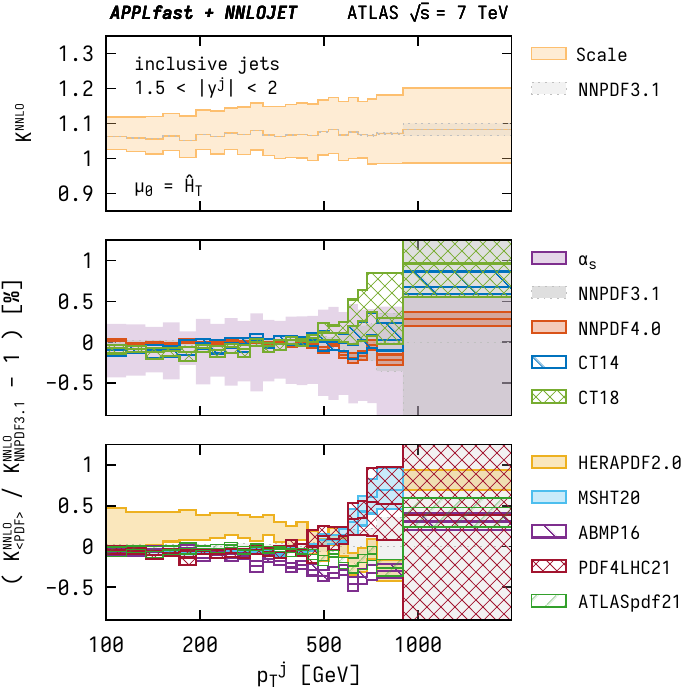}
  \caption{\label{fig:pdf_Kfac}%
    Summary of studies of the NNLO \kfactor for the ATLAS 7\,TeV inclusive jet cross section as a function of \ptjet for two representative ranges in rapidity.
    Top: the NNLO \kfactor and corresponding scale dependence.
    Middle and bottom panels: \kfactor for different PDF sets with their respective
    PDF uncertainty and dependence on the value of \asmz,
    shown as a ratio with respect to that obtained with NNPDF3.1\,.}
\end{figure*}

The na\"{i}ve application of a constant \kfactor using the reference scale $\muref$ as done in Eq.~\eqref{eq:NNLO_approx1}
gives rise to a scale uncertainty that is determined by the NLO component and thus at the $\pm5$--$10\%$ level for inclusive jets.
As a consequence, fits and extractions of SM parameters that are based on this approach and incorporate scale variations as
uncertainties will give rise to overly conservative estimates.  In cases where these uncertainties are sizeable,
such as \as extractions, a more reliable prediction is desirable.

The application in a scale-correlated manner as in Eq.~\eqref{eq:NNLO_approx2}, on the other hand, allows for
scale compensations to occur between $\sigma^{\NLO}$ and $K^\NNLO$, and it is necessary that the \kfactor be evaluated 
independently at both the central scale and all the other scales in question. This is, however, limited by the robustness of the
\kfactor with respect to changes of the PDFs, as the two terms in Eq.~\eqref{eq:NNLO_approx2} are in general evaluated
using different PDF parameterisations.  With the availability of NNLO grids, these assumptions can now be tested for
any PDF sets, including their full uncertainties.  Such a comparison is performed in Fig.~\ref{fig:pdf_Kfac} for 
two representative rapidity regions, where the top panel in each group shows $K^\NNLO(\mu)$ as defined 
in Eq.~\eqref{eq:Kfac}. The corresponding scale dependence is added for illustrative purpose and shown 
as the yellow filled bands that exhibit relative variations of $\pm5$--$10\%$.
While this scale variation can be associated with the ambiguity in the choice of $\muref$ in the na\"{i}ve 
approach~\eqref{eq:NNLO_approx1}, caution is advised in treating it as an additional uncertainty on top 
of the scale depende of $\sigma^{\NLO}(\mu)$ as it would lead to a substantial double-counting of the NLO-like scale variations.
In the case of the scale-correlated approach~\eqref{eq:NNLO_approx2}, it should be noted that the scale variations 
of $K^\NNLO(\mu)$ displayed here will cancel to a large extent with the correlated variation in $\sigma^{\NLO}(\mu)$, 
and therefore will not subsist in the final prediction.

The lower two panels in Fig.~\ref{fig:pdf_Kfac} display the relative difference between different NNLO $K$-factors for 
different PDF sets with respect to the baseline NNPDF3.1
prediction.  The middle panel further includes a shaded band (purple) that indicates the sensitivity to \as, again using   
variations in the range $0.108 \leq \asmz \leq 0.124$. 
Overall, a remarkable robustness of the NNLO \kfactor with respect to changes in the PDF set is observed at lower \pt,
agreeing to within the 0.5\% level.  The largest excursions are again seen for the ABMP16 and HERAPDF2.0 sets that,
nevertheless, still remain typically within $\pm0.5$--$1\%$.  The PDF uncertainties within a given PDF set are at a similar level.  
These results indicate that the approach following Eq.~\eqref{eq:NNLO_approx2} is, in general, 
likely to be safe also for fitting applications
where the \kfactors have been calculated independently for all of the required scales.
Nonetheless, the availability of interpolation grids frees us from having
to rely on such an assumption.

% -----------------------------------------------------------------------
\subsection{The total fiducial inclusive jet cross section}
\label{sec:incjet:totXS}
% -----------------------------------------------------------------------

The total fiducial inclusive jet cross section, $\sigma_\text{tot}^\text{jet}$, is
one of the largest inelastic cross sections measured in proton--proton collisions
and, as such, it is an important process for QCD studies.
Moreover, this process forms an important QCD induced background for many other processes measured at the LHC, 
so it is important to have a precise knowledge of the size of this cross section, and of the theoretical
uncertainties, in order to maximise the precision and physics potential of measurements from hadron--hadron colliders.

For this study, the total jet cross section is defined as the single-jet inclusive cross
section within a selected rapidity interval $|y^\text{jet}|$ and for
a minimal transverse momentum of \ptjetmin.
The ATLAS Collaboration has measured total jet cross sections for
anti-\kt jets with $R=0.4$ in the range $\ptjet>100\,$GeV and
$|y^\text{jet}|<3.0$ at centre-of-mass energies of 7, 8, and
13\,TeV~\cite{ATLAS:2014riz,ATLAS:2017kux,ATLAS:2017ble,ATLAS:2022djm}\footnote{The 
ATLAS Collaboration has also measured inclusive jet differential cross sections at $\sqrt{s}=2.76$ TeV~\cite{ATLAS:2013pbc}, 
but the bin boundaries chosen for that measurement preclude it from being used here.}. 
For the CMS Collaboration, the total jet cross section for centre-of-mass energies of
2.76, 7, 8, and 13\,TeV is derived from double-differential measurements for anti-\kt
jets with $R=0.7$~\cite{CMS:2015jdl,CMS:2014nvq,CMS:2016lna,CMS:2021yzl} by summing the cross sections in the bins of the common fiducial phase space of $\ptjet>97$\,GeV and $|y^\text{jet}|<2.0$.
The experimental uncertainties are obtained by propagating each uncertainty component individually and accounting for correlations.
The displayed total experimental uncertainty is then obtained by quadratic addition of all uncertainty components.

\begin{figure}
  \centering
  \includegraphics[width=\linewidth]{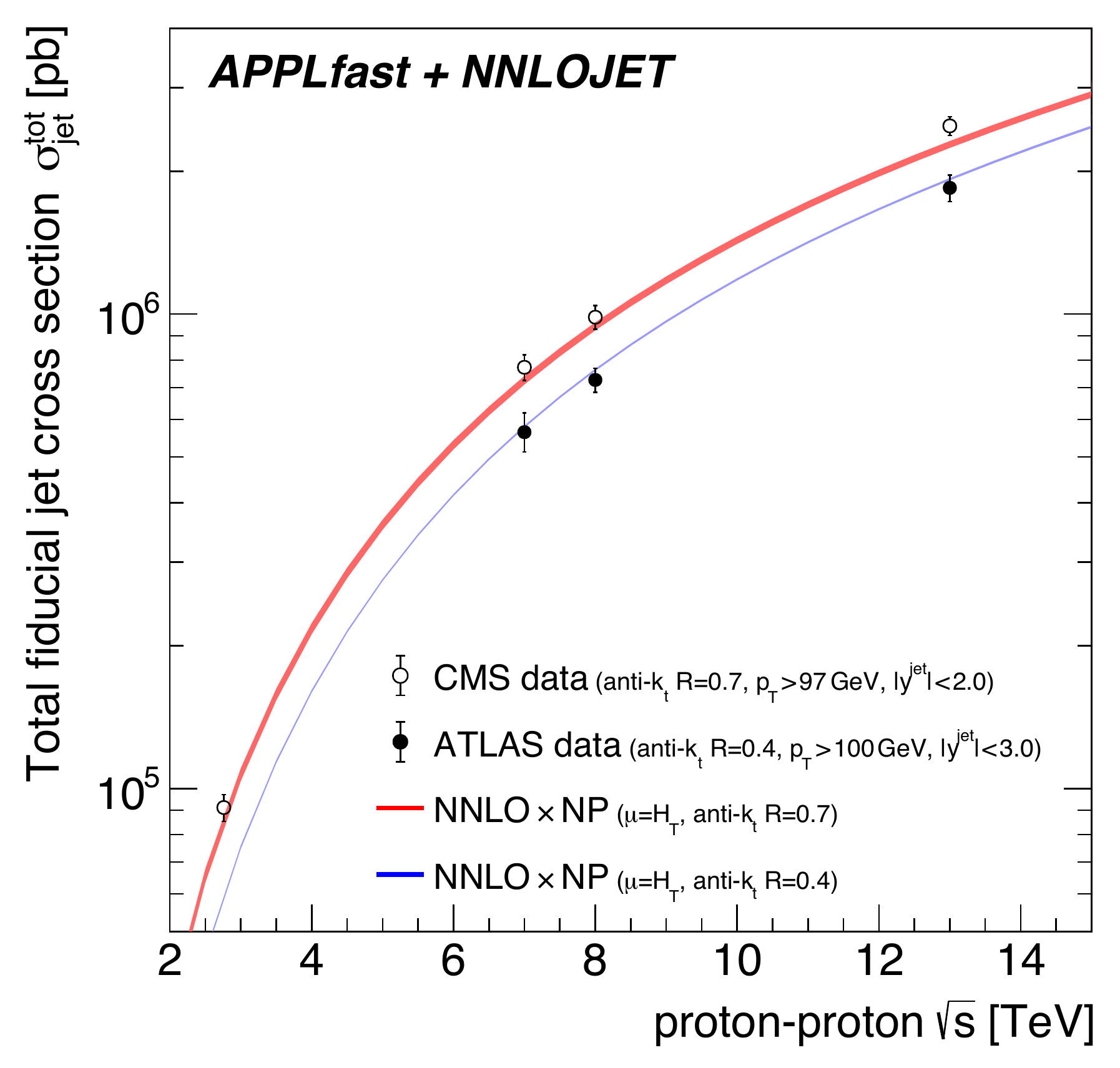}
  \caption{\label{fig:TotXsEcms}%
    The total jet cross section as a function of the proton--proton
    centre-of-mass energy for anti-\kt jets with $R=0.4$ and 0.7\,.
    The predictions are compared to data from ATLAS (for $R=0.4$) and
    CMS (for $R=0.7$), and the fiducial region is selected according
    to the available data.
    The size of the shaded area indicates the scale uncertainty, 
    evaluated as described in the text.
}
\end{figure}

Applying the technique of centre-of-mass reweighting~\cite{Carli:2010rw} for a grid, only the single grid at the largest
centre-of-mass energy of $\sqrt{s}=13$\,TeV is required for each jet-$R$ cone size.
For the NNLO predictions the PDF4LHC21 PDF set is used with the recommended scale of $\murf=\htp$.
Non-perturbative correction factors are taken from the relevant experimental publications as cross-section weighted averages and are
applied to the NNLO predictions.
Lacking bin-to-bin correlations, uncertainties for the non-perturbative corrections have not been derived.
Figure~\ref{fig:TotXsEcms} presents the results for the total jet cross section in comparison to data as a function of
the centre-of-mass energy for $R=0.4$ and $R=0.7$.  A reasonable agreement is observed between the data and the
predictions for all centre-of-mass energies.  The $\sqrt{s}$-dependence of the data is also well reproduced by the NNLO
predictions.

\begin{figure*}
  \centering
  \includegraphics[width=0.49\linewidth,trim={20 0 0 0},clip]{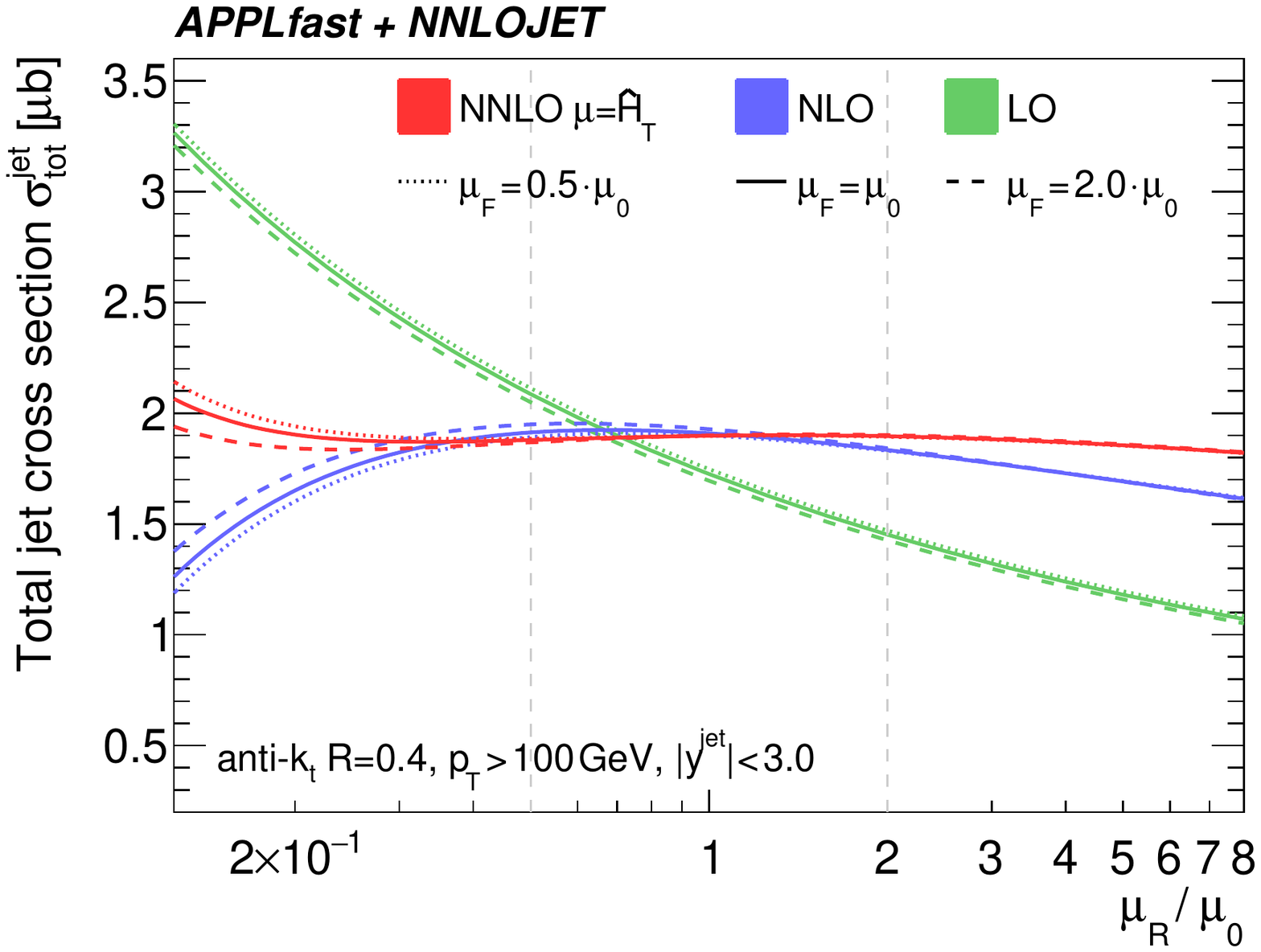}%
  \includegraphics[width=0.49\linewidth,trim={20 0 0 0},clip]{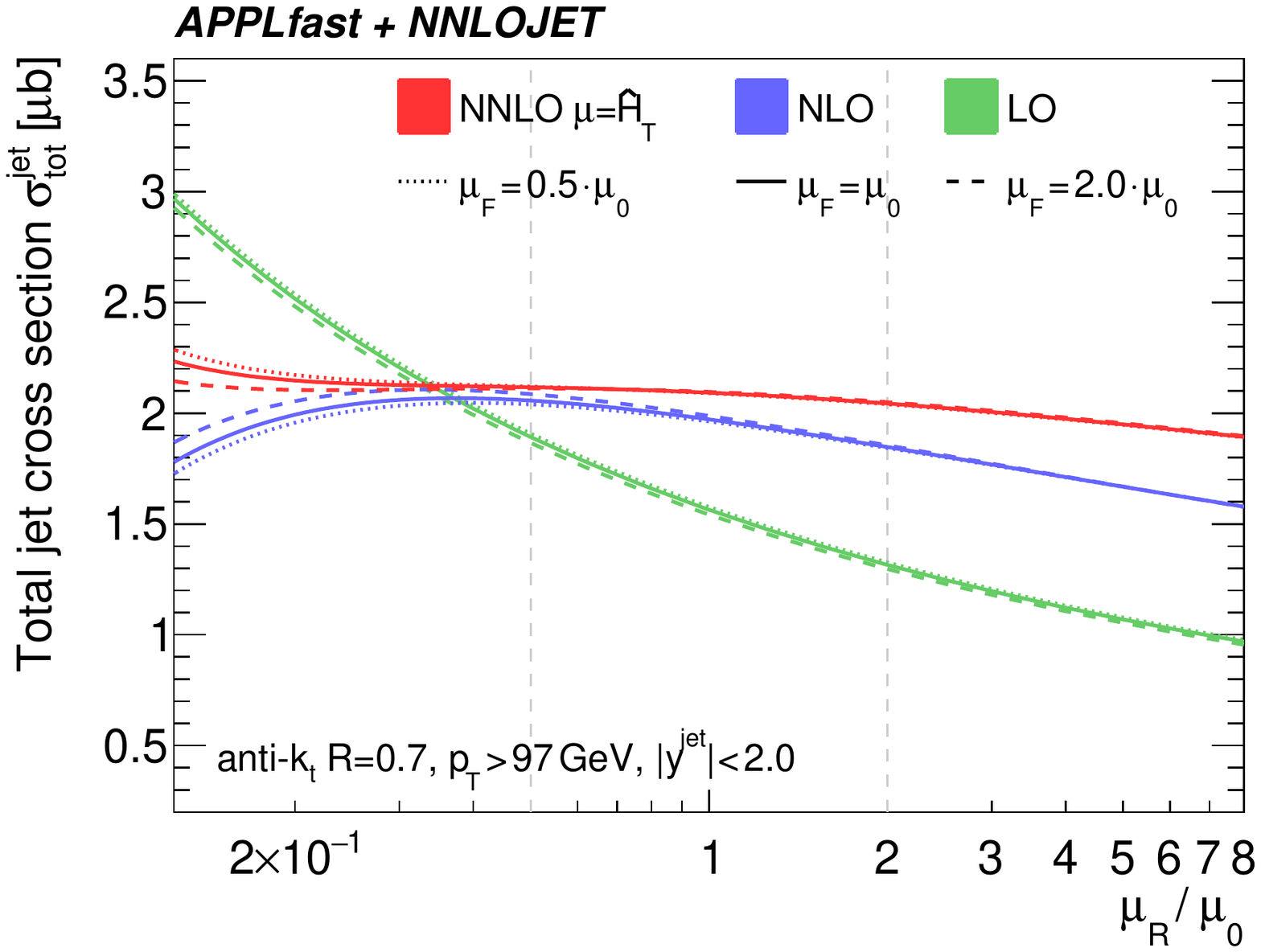}\\
  \includegraphics[width=0.49\linewidth,trim={20 0 0 0},clip]{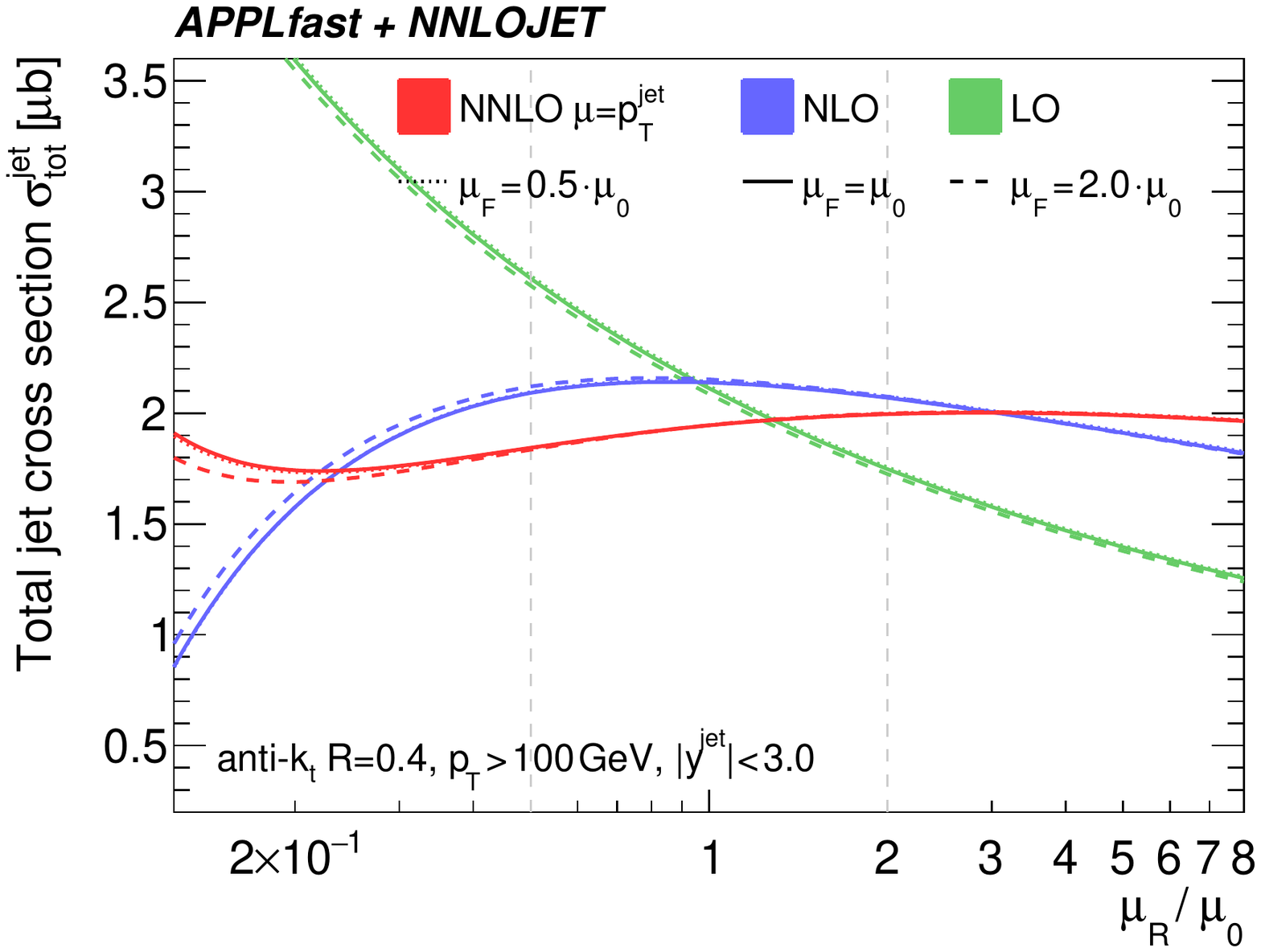}%
  \includegraphics[width=0.49\linewidth,trim={20 0 0 0},clip]{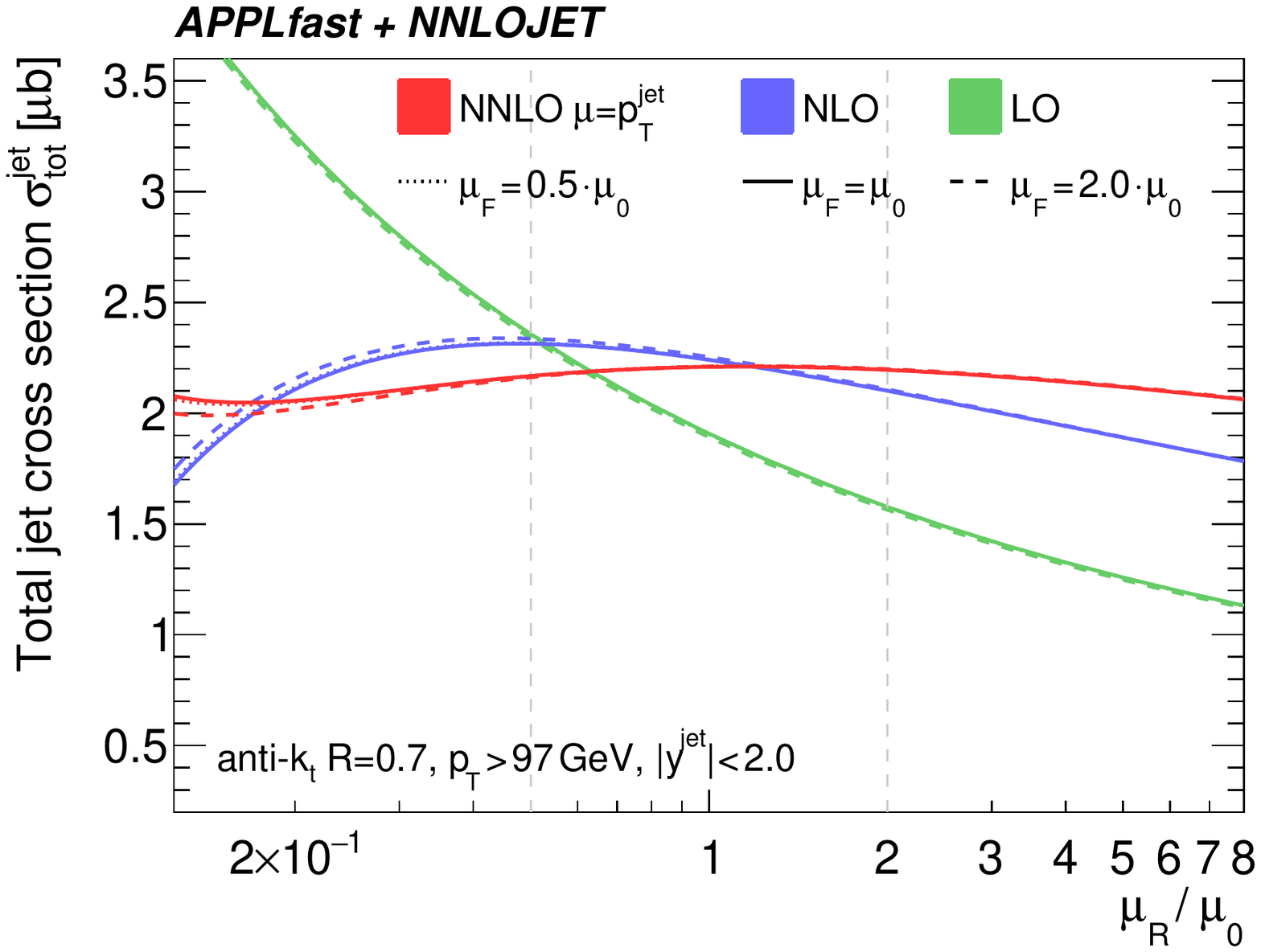}
  \caption{\label{fig:totXsScale}%
    The scale dependence of the total jet cross section at $\sqrt{s}=13$\,TeV for $R=0.4$ (left column) and $R=0.7$
    (right column) anti-\kt jets. The top row presents the scale dependence with $\murf = \htp$ as central scale, while
    the bottom row is for $\murf=\ptjet$.}
\end{figure*}

In the top row of Figure~\ref{fig:totXsScale} the scale dependence of $\sigma_\text{tot}^\text{jet}$ for $R=0.4$ and 0.7
at $\sqrt{s}=13$\,TeV is presented for $\murf=\htp$.  Shown is the total jet cross section at LO, NLO, and NNLO, where
scale factors ranging from 0.125 to 8 are applied to \mur, shown on the horizontal axis, and for the three scale factors
of 0.5, 1, and 2 for \muf shown in the band.  The NNLO correction is smaller than the NLO correction, and, as expected,
the scale dependence decreases moving from LO to NLO to NNLO. Both the NLO and the NNLO correction are somewhat smaller
for the cone size $R=0.4$ as compared to $R=0.7$.
For comparison, the scale dependence is also shown when using $\mu=\ptjet$ in the bottom row of
Figure~\ref{fig:totXsScale}.  In this case the NLO and the NNLO corrections decrease, but remain larger than unity for
the cone size of $R=0.7$, while they become very small at NLO and even smaller than unity at NNLO for $R=0.4$.
Moreover, the scale uncertainty at NNLO becomes larger than the one at NLO for the smaller cone size with $\mu=\ptjet$
confirming the findings of Ref.~\cite{Currie:2018xkj}.

Even though the predictions for the total jet cross section exhibit smaller scale uncertainties and smaller NNLO
\kfactors for $R=0.4$ than for $R=0.7$ for the recommended scale $\mu=\htp$, non-perturbative corrections have to be
considered as well for jet transverse momenta as small as $100\,$GeV.  For the cone radius $R=0.4$ the non-perturbative
correction is close to unity, while for $R=0.7$ it is of the order of 8\%~\cite{ATLAS:2017ble,CMS:2021yzl}. The
uncertainties on these corrections, however, are larger for the small cone size.

\begin{figure*}
  \includegraphics[width=0.48\linewidth,trim={60 0 0 0},clip]{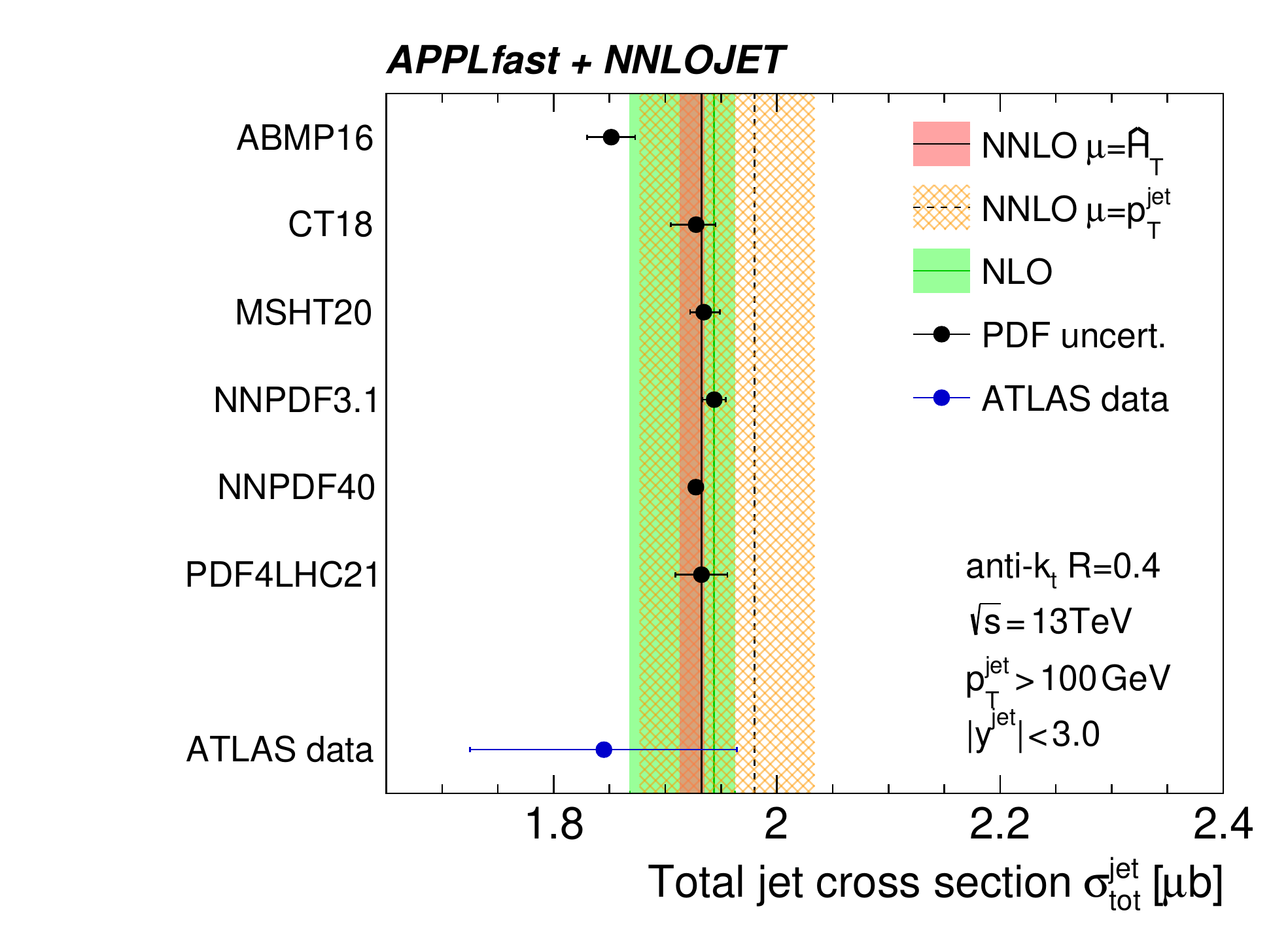}\hfill
  \includegraphics[width=0.48\linewidth,trim={60 0 0 0},clip]{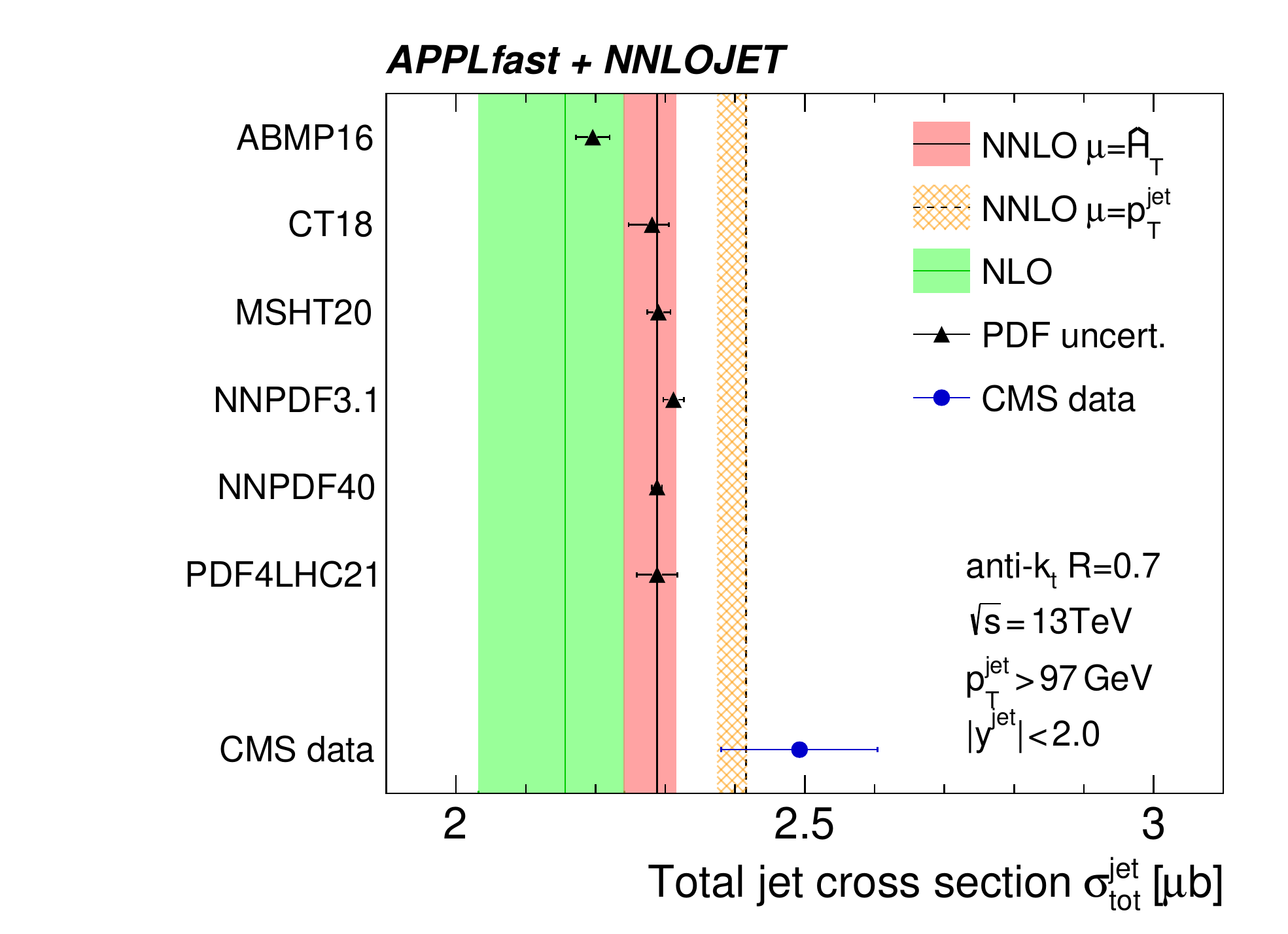}
  \caption{\label{fig:TotXsPDF}%
    The PDF dependence of the total jet cross sections for $R=0.4$ (left)
    and 0.7 (right) for the fiducial regions defined for ATLAS and CMS, respectively.
    Uncertainties of a few percent from non-perturbative corrections are not available
    for the total jet cross sections.}
\end{figure*}

The total jet cross section is also an important benchmark process for PDF
determinations, since it is sensitive to the gluon content of the proton.
In Figure~\ref{fig:TotXsPDF} the predicted total cross section $\sigma_\text{tot}^\text{jet}$
for various PDF sets (error bars indicating the respective PDF uncertainties only) is presented.
Without accounting for other theoretical uncertainties---in particular
those from non-perturbative corrections---it can be  observed that at $R=0.4$
the predictions are in agreement with the data, while for the larger $R=0.7$ radius 
the resulting cross section found using any of the PDF sets underestimates the measurement 
from CMS.

The predictions using the various PDF sets are
mutually compatible with the exception of the ABMP16 PDF set,
which predicts a significantly smaller cross section.
The PDF uncertainties exhibited by the different PDF sets vary in size
by approximately a factor of two. The recent NNPDF4.0 PDF set, however, estimates
PDF uncertainties to be significantly smaller than the others.

Also seen in Figure~\ref{fig:TotXsPDF} are the predictions at NLO and NNLO
with scale uncertainty bands. As expected from the discussion of
Figure~\ref{fig:totXsScale} the NNLO \kfactor is close to unity for
an anti-\kt jet radius of $R=0.4$, and larger for $R=0.7$.
Within the scale uncertainties, however, the predictions at NLO and NNLO remain
compatible with each other. In contrast, for $R=0.7$ the NNLO predictions
for the alternative scale of $\murf=\ptjet$ do not lie within the NLO
scale uncertainty.

% -----------------------------------------------------------------------

% -----------------------------------------------------------------------
%!TEX root = ../applfast.tex

% -----------------------------------------------------------------------
\section{PDF and \texorpdfstring{\as}{as} fits using dijet data}
\label{sec:dijet}
% -----------------------------------------------------------------------

In this section some of the full capabilities of the interpolation grids are illustrated
by performing a PDF fit using a single dijet
dataset from the measurements listed in Table~\ref{tab:dijet:datasets};
specifically, the CMS triple-differential dijet data at $\sqrt{s}=8\,\TeV$~\cite{CMS:2017jfq}.
Since this measurement has already been used in simultaneous PDF and \asmz fits at NLO QCD,
an example of those
results~\cite{Sieber2016_1000055723}, is first reproduced,
and then subsequently augmented by the inclusion of NNLO predictions.
The three dimensions of this dataset
divide the phase space into bins of the average transverse momentum of the two leading \pt jets,
$\ptave = (\ptoneb+\pttwob)/2$, the longitudinal boost of the dijet system given by half the sum of the leading jet
rapidities, $\yboost = |y_1+y_2|/2$, and half their rapidity separation, $\ystar = |y_1-y_2|/2$, which is related to the
scattering angle of the jets in the centre-of-mass system.

All PDF fits presented here are performed using \XFITTER\footnote{formerly known as \HERAFITTER.}~\cite{Alekhin:2014irh,Aaron_2010,Aaron_2009}
version 2.0.1, with technical updates required to fully exploit the new NNLO grids, as described in Ref.~\cite{Stark2020_1000131319}.
The details of the fits closely follow the HERAPDF2.0 methodology~\cite{H1:2015ubc},
with adaptions as described in Ref.~\cite{Sieber2016_1000055723}.
Further specific details of the fit parameterisation and procedure are described in the following subsections.

% ----- Dijet datasets
\begin{table*}[tbp]
  \caption{An overview of dijet datasets with \APPLFAST interpolation grids for proton--proton
    collisions at the LHC\@.
    For each dataset the centre-of-mass energy $\sqrt{s}$, the integrated luminosity $\mathcal{L}$,
    the number of data points, and the jet algorithm are listed.
    The two leading jets must fulfil the requirements with respect to their rapidities $y_1, y_2$ and
    transverse momenta \ptoneb, \pttwob. In addition, the choice of scale for \murf in the interpolation grids is shown.
  }
  \label{tab:dijet:datasets}
  \scriptsize\renewcommand{\arraystretch}{1.3}
  \begin{center}
    \begin{tabular}{l@{\hskip4pt}c@{\hskip4pt}c@{\hskip2pt}ccccc@{\hskip0pt}}
      \toprule
      \textbf{Data} & \textbf{\boldmath$\sqrt{s}$} & \textbf{\boldmath{$\mathcal{L}$}}
      & \textbf{no.\ of} & \textbf{anti-\boldmath{\kt}} & \textbf{kinematic range} & \textbf{fiducial cuts}
      & \textbf{\boldmath\murf-choice}
      \\
      & \textbf{[TeV]} & \textbf{\boldmath{$[{\rm fb}^{-1}]$}}
      & \textbf{points} & \textbf{\boldmath{$R$}} & \textbf{[GeV]} & &
      \\
      \midrule % arXiv:1312.3524, Inspire:1268975
      \multirow{3}{*}{ATLAS~\cite{ATLAS:2013jmu}} & %
      \multirow{3}{*}{7.0} & \multirow{3}{*}{4.5} & \multirow{3}{*}{90} &
      \multirow{3}{*}{$0.6$} & \multirow{3}{*}{$\mjj\in[260,5040]$} & $|y_1|,|y_2|<3.0$ & \multirow{3}{*}{$\mjj$}\\
                    &&&&&& $[\ptoneb,\pttwob]>[100,50]\GeV$&\\
                    &&&&&& $\ystar<3.0$&\\
      \midrule % arXiv:1212.6660, Inspire:1208923
      \multirow{3}{*}{CMS~\cite{CMS:2012ftr}} & %
      \multirow{3}{*}{7.0} & \multirow{3}{*}{5.0} & \multirow{3}{*}{54} &
      \multirow{3}{*}{$0.7$} & \multirow{3}{*}{$\mjj\in[197,5058]$} & $\yabs<5.0$ & \multirow{3}{*}{$\mjj$}\\
                    &&&&&& $[\ptoneb,\pttwob]>[60,30]\GeV$&\\
                    &&&&&& $|\ymax|<2.5$&\\
      \midrule % arXiv:1705.02628, Inspire:1598460
      \multirow{3}{*}{CMS~\cite{CMS:2017jfq}} &
      \multirow{3}{*}{8.0} & \multirow{3}{*}{$19.7$} & \multirow{3}{*}{122} &
      \multirow{3}{*}{$0.7$} & \multirow{3}{*}{$\ptave\in[133,1784]$} & $\yabs<5.0$ & $\ptmaxscl$\\
                    &&&&&& $\ptoneb,\pttwob>50\GeV$ &\\
                    &&&&&& $|y_1|,|y_2|<3.0$ & $\mjj$\\
      \midrule % arXiv:1711.02692, Inspire:1634970
      \multirow{4}{*}{ATLAS~\cite{ATLAS:2017ble}} & %
      \multirow{4}{*}{13.0} & \multirow{4}{*}{3.2} & \multirow{4}{*}{136} &
      \multirow{4}{*}{$0.4$} & \multirow{4}{*}{$\mjj\in[260,9066]$} & $|y_1|,|y_2|<3.0$ & \multirow{4}{*}{$\mjj$} \\
                    &&&&&& $\ptoneb,\pttwob>75\GeV$&\\
                    &&&&&& $\ptave>100\GeV$&\\
                    &&&&&& $\ystar<3.0$&\\
      \bottomrule
    \end{tabular}
  \end{center}
\end{table*}

\subsection{Reproduction of previous fits at NLO}
\label{sec:dijet:reproduction}

In order to validate the PDF fitting procedure used for this analysis, two NLO fits are first performed:
one using the HERA I+II inclusive DIS data alone, and another that additionally includes
the CMS triple-differential dijet data.

As prescribed by the HERAPDF2.0 procedure, the PDFs $f_i(x)$ are parameterised at some starting scale \mufn by
\begin{equation}
  xf_i(x) = A_i x^{B_i} (1-x)^{C_i} (1+{D_i}x+{E_i}x^2)\,,
  \label{eqn:dijet:pdfgen}
\end{equation}
where the parameters $A_i$, $B_i$, and $C_i$ are always included, while the $D_i$ and $E_i$ parameters
increase the flexibility of the fit and can be used to estimate the parameterisation uncertainties.
To describe the proton, five such PDFs are parameterised,
defined here to be: the gluon $f_\text{g}$; the valence quarks $f_{\text{u}_\text{v}}=f_\text{u}-f_{\overline{\text{u}}}$ and
$f_{\text{d}_\text{v}}=f_\text{d}-f_{\overline{\text{d}}}$; and the light up- and down-type anti-quark distributions
$f_{\overline{\text{U}}}=f_{\overline{\text{u}}}$ and $f_{\overline{\text{D}}}=f_{\overline{\text{d}}}+f_{\overline{\text{s}}}$.
It should be noted that the default HERAPDF2.0 parameterisation, also used in Ref.~\cite{CMS:2017jfq},
includes a second subtracted term of the form $A'_\text{g} x^{B'_\text{g}} \left(1-x\right)^{C'_\text{g}}$ for the gluon distribution,
for fits beyond LO. Following Ref.~\cite{Sieber2016_1000055723}, this term is not adopted here, since it offers
no advantage in \chisqndof for the performed studies.

Of the five normalisation constants $A_i$, three are constrained by the quark-number and momentum sum rules.
Following HERAPDF2.0 choices, a symmetric low-$x$ behaviour of the up-and down-type quark sea is assumed,
the strange sea distribution is written as a fixed fraction $f_{\overline{\text{s}}/\overline{\text{D}}}=0.4$ of the down-type quark
sea\footnote{Reference~\cite{H1:2015ubc}
labels this parameter simply as $f_s$, which differs from the PDF notation adopted here.}, and it is assumed that
$x\text{s}=x\overline{\text{s}}$.
Finally, the $\overline{\text{u}}$ and $\overline{\text{d}}$ anti-quark normalisations are constrained
to be equal in the limit $x\rightarrow 0$.
This would leave ten free parameters if all $D_i$ and $E_i$ were set to zero.

Following Ref.~\cite{Sieber2016_1000055723}, specific differences with respect to the published
HERAPDF2.0\footnote{The default HERAPDF2.0, as published, uses $Q^2_\textrm{min}=3.5\,\GeV^2$, and a 14-parameter fit: 10
$+\, D_{\overline{\text{U}}}$, $E_{\text{u}_\text{v}}$, $A'_{\text{g}}$ and $B'_{\text{g}}$.}
are: a larger minimum $Q^2$ cut for the DIS data of $Q^2_\textrm{min}=7.5\,\GeV^2$,
the non-inclusion of the negative gluon term, and a choice of a 13-parameter fit at NLO, with the parameters $E_\text{g}$,
$D_{\text{u}_\text{v}}$ and $D_{\overline{\text{U}}}$ included, as these were found to optimally fit the CMS
triple-differential dijet data when added to the minimal set of ten parameters.
Additional differences with respect to the PDF fit described in the CMS publication~\cite{CMS:2017jfq}
are summarised in~\ref{sec:dijetfits}.
The theoretical calculation used to fit the CMS dijet measurement is from \NLOJETPP~\cite{Nagy:2001fj,Nagy:2003tz},
encoded in the fast interpolation grids of \FASTNLO.

The starting values of the parameters for the fit to HERA DIS data alone are set to those published by the HERAPDF2.0 analysis~\cite{H1:2015ubc}, except for the parameters $E_\text{g}$ and $D_{\text{u}_\text{v}}$, which were not fitted there and are given a starting value of 0.
The values of all 13 parameters resulting from this fit are then used as starting values for the simultaneous fit to the HERA DIS and the CMS dijet data.

Following Ref.~\cite{Sieber2016_1000055723}\footnote{At the time of Ref.~\cite{Sieber2016_1000055723},                                       
electroweak corrections were not available, and so the CMS dijet data were restricted to a phase space region                                                                where these were expected to be small -- see \ref{sec:dijetfits}}, 
the CMS dijet data used were
limited to the range \mbox{$\ptave < 1\,\TeV$}
and without electroweak corrections.
The results are in good agreement with those in Ref.~\cite{Sieber2016_1000055723}.
Since electroweak corrections are now available, the NNLO fit presented in Section~\ref{sec:dijet:nnloextension}
includes these, and uses the full $\ptave$ range of the measurement.
For validation, an additional fit at NLO was performed including the CMS $\ptave$
data beyond $1\,$TeV, with electroweak corrections applied, and taking into account two additional uncertainty sources,
as described in \ref{sec:dijetfits}.
It was found that these modifications lead to only negligible differences, supporting the contention that their impact is limited,
due to the larger statistical uncertainties on the jet data at high scales, and the small size of the electroweak corrections at smaller
scales. 
Replacing the NLO prediction from \NLOJETPP with that from \NNLOJET leaves the results practically unchanged,
as expected.
Further details can be found in Ref.~\cite{Stark2020_1000131319}.

\subsection{Extension to NNLO}
\label{sec:dijet:nnloextension}

Following the initial validation of the fit procedure to reproduce the previous fit, the methodology was
extended to include the NNLO predictions available from \NNLOJET. The corresponding
interpolation grids have been created with two different central scale choices: $\murf=\ptmaxscl$ as before, and
$\murf=\mjj$, the mass of the dijet system, as recommended in Ref.~\cite{Gehrmann-DeRidder:2019ibf}. To be less
sensitive to potential issues of the theoretical description at low $x$, such as
the need for resummation corrections~\cite{Ball_2018} or the impact of higher-twist corrections at low $Q^2$,
the minimum $Q^2$ for the DIS data is increased to $Q^2_\textrm{min}=10.0\,\GeV^2$.

\begin{figure*}
  \includegraphics[width=0.48\linewidth]{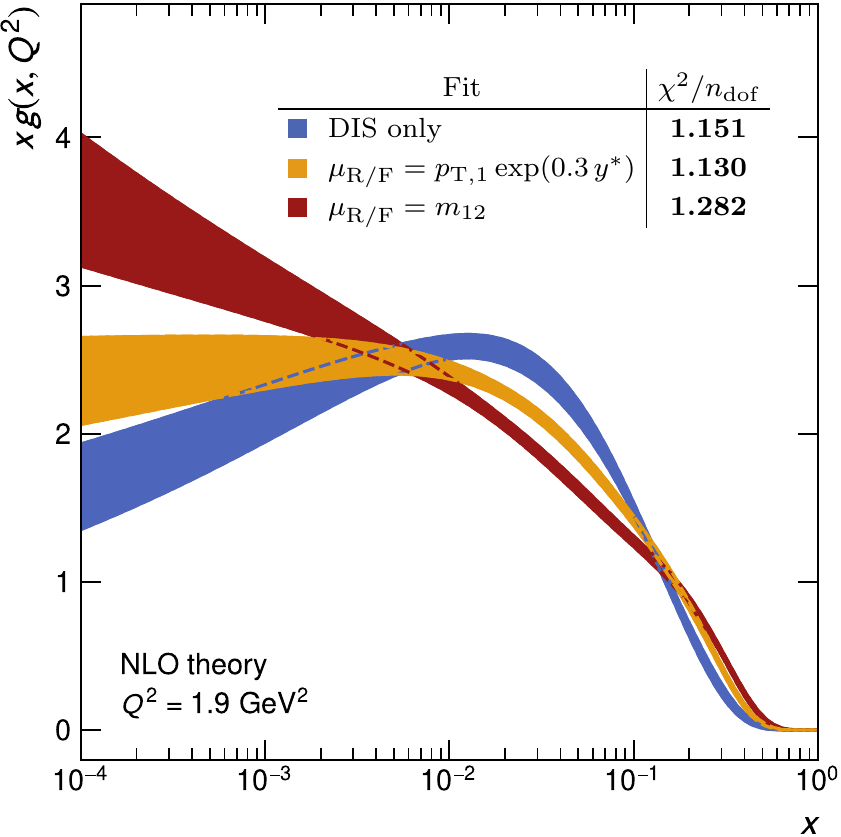}\hfill
  \includegraphics[width=0.48\linewidth]{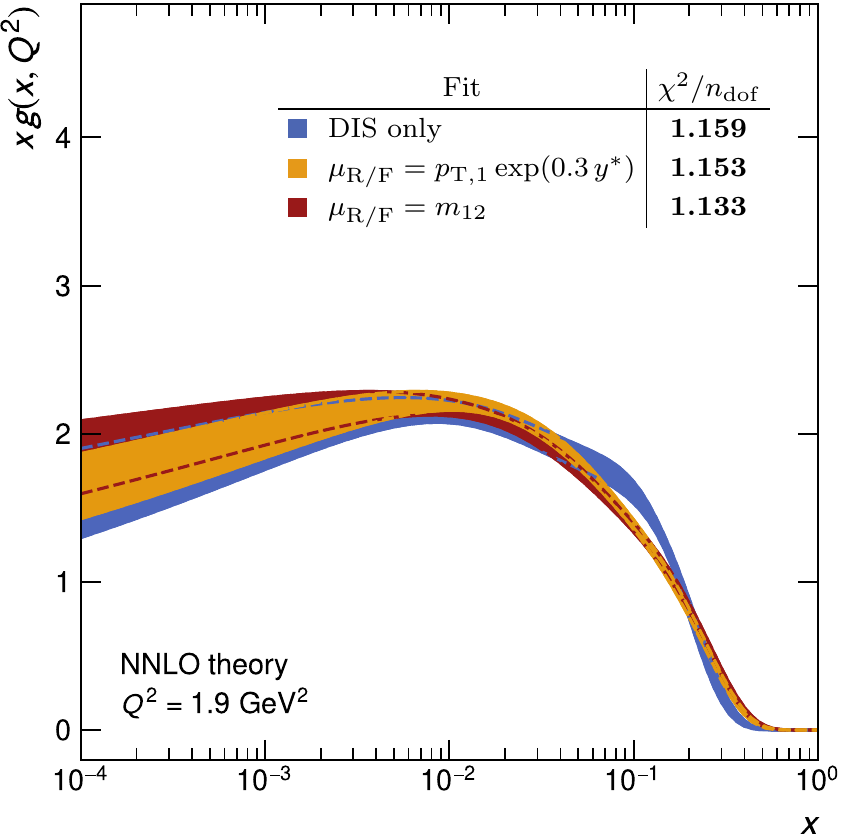}
  \caption{The gluon PDF from fits to HERA DIS alone
    and with CMS dijet data at NLO (left) and NNLO (right) for two different central scale choices for dijet production:
    $\murf=\ptmaxscl$ (yellow)
    and $\murf=\mjj$ (red).
    Only the experimental uncertainties are shown.
    }
  \label{dijet:fig:pdffits}
\end{figure*}

By examining
the gluon distribution -- the most sensitive parton distribution --  the dependence of the fit result on the central scale choice
is investigated, as shown in Fig.~\ref{dijet:fig:pdffits}. At NLO (left), significant differences are observed for the two scales,
while at NNLO (right) all fit qualities improve and
the differences resulting from the two scales are drastically
reduced such that the uncertainty bands overlap even though they represent only the experimental uncertainty.
It is perhaps interesting to note that
the recommended scale  $\murf=\mjj$ exhibits, at NLO, a significantly worse \chisqndof of~$1.282$ than the value $1.130$
obtained using the scale \ptmaxscl, while at NNLO the scale choice of \mjj exhibits
the best value of \chisqndof, close to unity.

\begin{figure*}
  \includegraphics[width=0.48\linewidth]{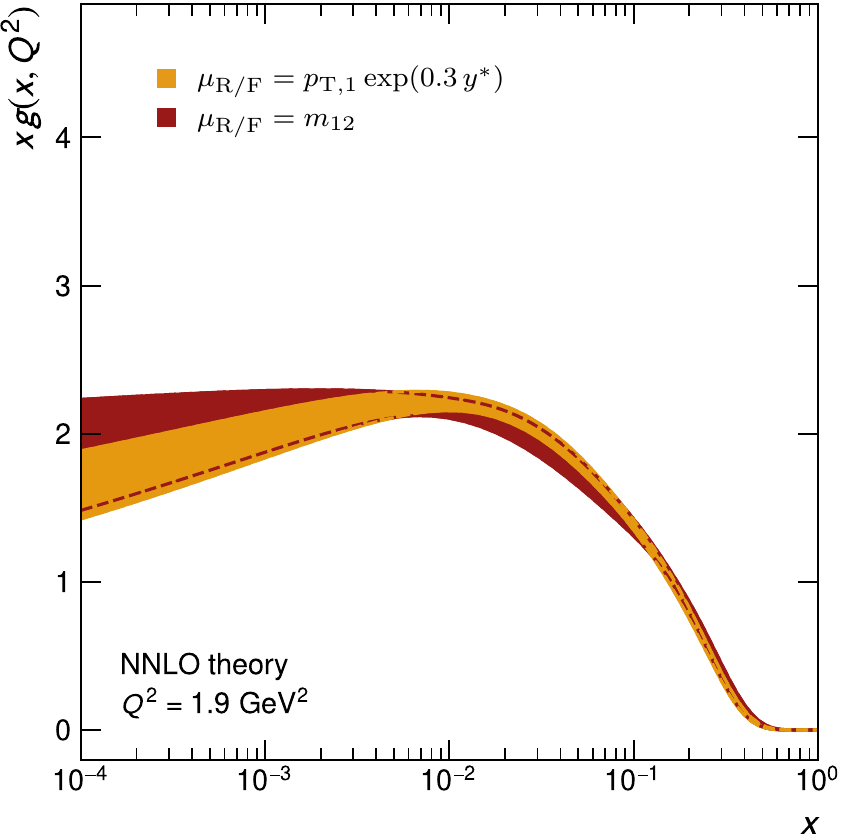}\hfill
  \includegraphics[width=0.48\linewidth]{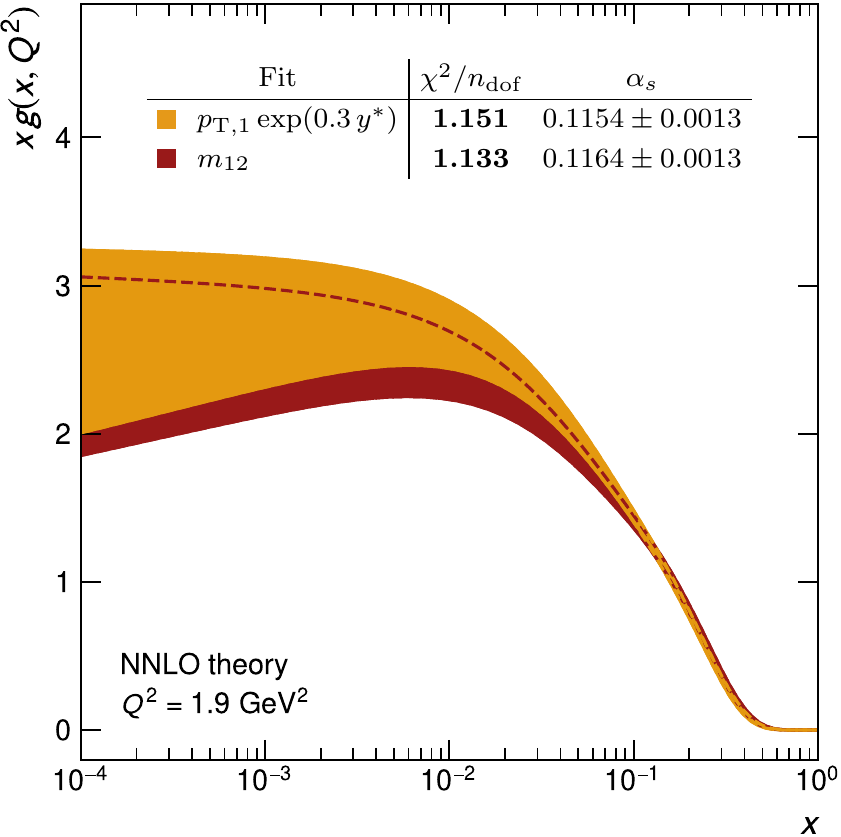}
  \caption{Fits at NNLO of the gluon PDF using HERA DIS and CMS dijet data, but with uncertainties including scale
    variations (left) as described in the text, or with \asmz as an additional parameter (right).
  }
  \label{dijet:fig:pdffitsscaleas}
\end{figure*}

Employing the conventional scale variation methodology
as a proxy for the effect of missing higher orders,
the fits for both scales overlap as demonstrated in Fig.~\ref{dijet:fig:pdffitsscaleas} (left),
where the scale variations are considered as an
additional uncertainty in the form of an envelope (evaluated using the ``offset method''),
added quadratically to the experimental uncertainty.
Detailed fit results can be found in Table~\ref{app:tab:pdffitsnnlocomb} in~\ref{sec:dijetfits},
and in Ref.~\cite{Stark2020_1000131319}, where also a simultaneous fit to both CMS dijet
datasets of Table~\ref{tab:dijet:datasets} is discussed.

\begin{table*}[tbp]
  \caption{Values of \asmz determined in the (13+1)-parameter PDF+\asmz fit at NLO and at NNLO with HERA DIS and CMS triple-differential
    dijet data, as described in the text. 
  }
  \label{dijet:tab:alphas8tev}
  \scriptsize \renewcommand{\arraystretch}{1.3}
  \begin{center}
    \begin{tabular}{llcccccc}
      \toprule
      \multirow{2}{*}{order} & scale      & \multirow{2}{*}{\asmz} & \multicolumn{5}{c}{$\Delta\asmz\cdot10^4$} \\
                             & choice     &                        & (exp)       & (scale)                            & (model)      & (param)      & (total) \\
      \midrule
      \multirow{2}{*}{NLO}   & \ptmaxscl  & 0.1192                 & $\pm 15$    & $^{+26}_{-\phantom{0}9}$           & $^{+5}_{-4}$ & $^{+1}_{-3}$ & $^{+31}_{-19}$\\
                             & \mjj       & 0.1210                 & $\pm 15$    & $^{+32}_{-27}$                     & $\pm 4$      & $^{+6}_{-7}$ & $^{+36}_{-32}$\\
      \midrule
      \multirow{2}{*}{NNLO}  & \ptmaxscl  & 0.1154                 & $\pm 13$    & $^{+\phantom{0}7}_{-\phantom{0}8}$ & $^{+5}_{-4}$ & $^{+1}_{-7}$ & $^{+15}_{-17}$\\
                             & \mjj       & 0.1164                 & $\pm 13$    & $^{+11}_{-\phantom{0}5}$           & $^{+6}_{-4}$ & $^{+2}_{-6}$ & $^{+18}_{-16}$\\
      \bottomrule
    \end{tabular}

  \end{center}
\end{table*}

\subsection{Full fit of PDFs and \asmz}
\label{sec:dijet:fullfit}

The procedure established in Section~\ref{sec:dijet:nnloextension} is extended further by including \asmz as an
additional free parameter in the fit.  This leads to similar results for the PDFs as before, but with larger uncertainties, in
particular for the gluon, shown in Fig.~\ref{dijet:fig:pdffitsscaleas} (right).

Table~\ref{dijet:tab:alphas8tev} shows the values of \asmz obtained for the simultaneous PDF and \asmz fits. It is
observed that the values preferred by the NNLO fits are smaller than those obtained at NLO.
The fit quality, compared in the inserts in Figs.~\ref{dijet:fig:pdffits} (right) and~\ref{dijet:fig:pdffitsscaleas} (right), is largely unchanged,
since the previously fixed value of $\asmz=0.1180$ is already close to the minimum found here.  The experimental
uncertainty on the quoted value of \asmz is determined from the parabolic dependence of the $\chi^2$ function near the minimum,
and the scale uncertainty is obtained as described in Section~\ref{sec:dijet:nnloextension}, using the offset method.

Two additional sets of uncertainties are estimated, to account for the chosen PDF parameterisation and the values of certain
model and procedural choices, the latter detailed in Table~\ref{app:tab:modelpars} of~\ref{sec:dijetfits}.
The parameterisation uncertainty is determined by including all the remaining $D$ and $E$ parameters in the fit one
at a time and taking the largest positive and negative difference of any of these variations,
with respect to the nominal \asmz value, as an asymmetric
uncertainty.  To obtain the model uncertainty, fits are performed for up and down variations of the masses of the charm quark, $m_\text{c}$,
and bottom quark, $m_\text{b}$, the strangeness fraction, $f_{\overline{\text{s}}/\overline{\text{D}}}$,
the starting scale, $\mu_{\text{F}_0}$, and the $Q^2_\textrm{min}$ cut imposed on the DIS data.
Table~\ref{app:tab:modelpars} lists the values of the parameters used in the nominal fit and for the systematic
variations. The signed differences with respect to the nominal \asmz value, resulting from the variation of each parameter,
are added in quadrature to yield the overall model uncertainty. The final result using the recommended scale choice
of $\murf=\mjj$ is $\asmz = 0.1164\,^{+0.0018}_{-0.0016}\,\mathrm{(tot)}$,
which is compatible with the result using the alternative scale choice (see Table~\ref{dijet:tab:alphas8tev})
and the world average of $\asmz = 0.1179 \pm 0.0009$~\cite{ParticleDataGroup:2020ssz}.

% -----------------------------------------------------------------------

% -----------------------------------------------------------------------
%!TEX root = ../applfast.tex

% -----------------------------------------------------------------------
\section{Conclusions and outlook}
\label{conclusions}
% -----------------------------------------------------------------------

The technique of interpolation grids has been proven to be an indispensable tool for QCD phenomenology of hadron-collider
data, since it allows repeated calculations of pQCD cross sections with varying input conditions such as scale choices,
parton distribution functions, or the strong coupling, \as. The grid technique is implemented in the
\APPLGRID and \FASTNLO computer codes and, for this paper, a common interface for both packages with the
\NNLOJET computer program for hadron--hadron processes has been developed, enabling the generation of fast interpolation grids
at next-to-next-to-leading order in QCD for inclusive jet
and dijet production cross sections at the LHC\@.

The performance of the grid technique for selected LHC jet datasets is presented, demonstrating a closure of
those interpolation grids generated for inclusive jet and dijet cross sections to generally better than 0.1\%.
Although the NNLO calculations for jet production are computationally very expensive, all grids were
generated such that the statistical uncertainty is less than 1\% in most cases, but may increase to
2--4\% for some regions near the edges of the phase space from the measurement.

The grids have been employed for phenomenological studies that are otherwise computationally prohibitive. 
The NNLO predictions for inclusive jet and dijet cross sections were evaluated for different PDF sets including
the full PDF uncertainties.  It is found
that the PDFs from the CT, MSHT, and NNPDF global fitting groups yield largely consistent predictions for jet production
cross sections over a large kinematic range in $\ptjet$ and jet rapidity $y$. The predictions using the ABMP16 or
HERAPDF2.0 PDFs exhibit some deviations beyond the PDF uncertainty bands, e.g.\ at large $\ptjet$ or $y$.  The new
NNPDF4.0 set yields surprisingly small PDF uncertainties compared to the other available PDF sets.

Predictions for NNLO cross sections are often included in QCD phenomenological studies through the use of NNLO \kfactors.
Here, their stability while varying the strong coupling \as or the PDF set has been studied.
Overall, it is observed that the NNLO \kfactors are largely insensitive to the choice of \as as expected, but exhibit a dependence on the PDF sets at the few permille level in the bulk of phase space that increases to the percent level towards the tails of the \pt distributions.
The latter are sub-dominant compared to NNLO scale uncertainties, though not necessarily negligible.
Therefore, although a case could be made to not require the full grid for reproduction of the central cross section, for
full precision and, in particular, for the evaluation of the scale uncertainties, the use of interpolation 
grids is preferred over a \kfactor approach.

The extension of the grid framework to hadron--hadron collider processes shown here further enables the generation of
interpolation grids for the large set of processes available within \NNLOJET. Such applications and phenomenological 
studies will be left for future studies. 

The grids used in this analysis correspond to a large number of the available jet measurements 
from the ATLAS and CMS collaborations. They have been made available for the wider community 
on the \PLOUGHSHARE~\cite{web:ploughshare} website.

% -----------------------------------------------------------------------

% -----------------------------------------------------------------------
%!TEX root = ../applfast.tex

% -----------------------------------------------------------------------
\section*{Acknowledgements}
% -----------------------------------------------------------------------

We thank J.~Hessler for producing early versions of the dijet grids.
This research was supported in part by the UK Science and Technology Facilities Council (STFC) through grant ST/T001011/1, by the Swiss National Science
Foundation (SNF) under contracts 200021-197130 and 200020-204200, by the Research Executive Agency (REA) of the European
Union through the ERC Advanced Grant MC@NNLO (340983), and by the state of Baden-Württemberg through
bwHPC and the German Research Foundation (DFG) through grant no INST 39/963-1 FUGG (bwForCluster NEMO).  CG and MS were
supported by the IPPP Associateship program for this project.  JP was supported by
Funda\c{c}\~ao para a Ci\^encia e Tecnologia (FCT, Portugal) through the project funding of R\&D units reference UIDP/50007/2020 and under project CERN/FIS-PAR/0024/2019.

% -----------------------------------------------------------------------

% -----------------------------------------------------------------------
\onecolumn
\appendix
% -----------------------------------------------------------------------

% -----------------------------------------------------------------------
%!TEX root = ../applfast.tex

% -----------------------------------------------------------------------
\section{Details on the dijet fits}
\label{sec:dijetfits}
% -----------------------------------------------------------------------

Additional setup differences in the reproduction fits of Section~\ref{sec:dijet:reproduction} with respect to the PDF
fit described in the CMS publication Ref.~\cite{CMS:2017jfq} are:

\begin{enumerate}
\item A newer version of the fitting framework, \XFITTER v2.0.1, is used in preference to \HERAFITTER v1.1.1.
\item The parameter fixing the strange sea fraction, $f_{\overline{\text{s}}/\overline{\text{D}}}$, is set to $0.4$
  instead of $0.31$.
\item At the time of writing of the PhD thesis Ref.~\cite{Sieber2016_1000055723} electroweak correction factors were not
  available. The range in \ptave for fits hence was restricted to $\ptave < 1\,\TeV$.
\item The data comprise an additional experimental source of uncertainty that accounts for non-Gaussian tails
  in the jet energy resolution.
\item The statistical and numerical precision of the theory calculations is taken into account as an
  additional bin-to-bin uncorrelated uncertainty in the fit.
\item New dijet predictions at NLO pQCD have been calculated with interpolation grids produced with
  \NNLOJET~\cite{Currie:2017eqf} instead of \NLOJETPP~\cite{Nagy:2001fj,Nagy:2003tz}.
\end{enumerate}

The detailed results of the dijet fits of Section~\ref{sec:dijet:nnloextension} are presented here in
Tables~\ref{app:tab:pdffitsnlo}--\ref{app:tab:pdffitsnnlocomb}. The labels ``yb0 ys0'' up to ``yb2 ys0'' denote the six
measurement bins in \yboost and \ystar with $0,1,2$ corresponding to $0 < \yboost|\ystar$, $1 \leq \yboost|\ystar < 2$,
and $2 \leq \yboost|\ystar < 3$.

The variations used for the model uncertainties are summarised in Table~\ref{app:tab:modelpars}.

\begin{table*}[htb]
  \caption{Partial $\chi^2$ values for the NLO fits of Section~\ref{sec:dijet:nnloextension}}
  \label{app:tab:pdffitsnlo}
  \centerline{\small
    \begin{tabular}{rlc|ccc}
      & & \multirow{2}{*}{$n_\mathrm{data}$} & HERA I+II & with CMS dijets & with CMS dijets \\
      & & & DIS only & $\murf=\ptmaxscl$ & $\murf=\mjj$ \\\hline
      \multirow{1}{*}{HERA I+II}
      & combined & 1016 & 1106.14 & 1124.45 & 1157.94 \\\hline
      \multirow{7}{*}{CMS $8\,\TeV$ dijets}
      & yb0 ys0  &   31 &  -- &   14.58 &   29.13 \\
      & yb0 ys1  &   26 &  -- &   11.36 &   22.41 \\
      & yb0 ys2  &   14 &  -- &   17.77 &   52.60 \\
      & yb1 ys0  &   23 &  -- &   11.29 &   20.41 \\
      & yb1 ys1  &   17 &  -- &   18.87 &   21.69 \\
      & yb2 ys0  &   11 &  -- &   19.51 &   58.30 \\\cdashline{2-6}
      & combined &  122 &  -- &   93.38 &  204.54 \\\hline
      \multicolumn{3}{r|}{correlated $\chi^2$}  &   50.96 &   65.04 &   92.11 \\
      \multicolumn{3}{r|}{log penalty $\chi^2$} &   -2.98 &  -11.39 &  -12.89 \\\hline
      \multicolumn{3}{r|}{combined}             & 1154.12 & 1271.48 & 1441.72 \\
      \multicolumn{3}{r|}{\ndof}                &    1003 &    1125 &    1125 \\
      \multicolumn{3}{r|}{p-value}              & \num{6.13d-4} & \num{1.45d-3} & \num{3.86d-10} \\
      \multicolumn{3}{r|}{combined \chisqndof}  & \bf 1.151 & \bf 1.130 & \bf 1.282 \\
    \end{tabular}
  }
  \vspace*{0.5cm}
\end{table*}

\begin{table*}[htb]
  \caption{Partial $\chi^2$ values for the NNLO fits of Section~\ref{sec:dijet:nnloextension}}
  \label{app:tab:pdffitsnnlocomb}
  \centerline{\small
    \begin{tabular}{rlc|ccc}
      & & \multirow{2}{*}{$n_\mathrm{data}$} & HERA I+II & with CMS dijets & with CMS dijets \\
      & & & DIS only & $\murf=\ptmaxscl$ & $\murf=\mjj$ \\\hline
      \multirow{1}{*}{HERA I+II}
      & combined & 1016 & 1109.15 & 1115.65 & 1124.45 \\\hline
      \multirow{7}{*}{CMS $8\,\TeV$ dijets}
      & yb0 ys0  &   31 &      -- &   18.15 &   14.58 \\
      & yb0 ys1  &   26 &      -- &   13.07 &   11.36 \\
      & yb0 ys2  &   14 &      -- &   26.63 &   17.77 \\
      & yb1 ys0  &   23 &      -- &   14.45 &   11.29 \\
      & yb1 ys1  &   17 &      -- &   23.91 &   18.87 \\
      & yb2 ys0  &   11 &      -- &   18.19 &   19.51 \\\cdashline{2-6}
      & combined &  122 &      -- &  114.40 &   93.38 \\\hline
      \multicolumn{3}{r|}{correlated $\chi^2$}   &         55.48 &         65.78 &         61.47 \\
      \multicolumn{3}{r|}{log penalty $\chi^2$}  &         -1.74 &          1.45 &         -1.67 \\\hline
      \multicolumn{3}{r|}{combined}              &       1162.89 &       1297.30 &       1274.63 \\
      \multicolumn{3}{r|}{\ndof}                 &          1003 &          1125 &          1125 \\
      \multicolumn{3}{r|}{p-value}               & \num{3.24d-4} & \num{2.55d-4} & \num{1.19d-3} \\
      \multicolumn{3}{r|}{combined \chisqndof}   & \bf 1.159 & \bf 1.153 & \bf 1.133 \\
    \end{tabular}
  }
  \vspace*{0.5cm}
\end{table*}

\begin{table*}[htb]
  \caption{Values of the model parameters used in the fits in Sections~\ref{sec:dijet:nnloextension}
    and~\ref{sec:dijet:fullfit}, and the corresponding systematic variations used to determine the model uncertainty on
    \asmz in Section~\ref{sec:dijet:fullfit}. Omitted values are due to the restriction $\mu_{\text{F}_0} < m_\text{c}$
    imposed by the RTOPT heavy flavour number scheme used here~\cite{Thorne:1997ga, Thorne:2006qt, Thorne:2012az}. In
    these cases, to calculate the corresponding contributions to the model uncertainty, the symmetrised variation in the
    opposite direction is taken.}
  \label{app:tab:modelpars}
  \centerline{\small
    \begin{tabular}{c|ccc}
      \multirow{2}{*}{Parameter}     & Central  &  \multicolumn{2}{c}{Variation} \\
                                     & value    &  down  &  up  \\\hline
      $f_{\overline{\text{s}}/\overline{\text{D}}}$ & 0.4      &  0.3   &  0.5 \\
      $m_\text{c}$ (GeV)             & 1.42     &  --    &  1.49 \\
      $m_\text{b}$ (GeV)             & 4.5      &  4.25  &  4.75 \\
      $\mu_{\text{F}_0}^2$ (GeV$^2$) & 1.9      &  1.6   &  --  \\
      $Q^2_\textrm{min}$ (GeV$^2$)   & 10       &  7.5   &  12.5 \\
    \end{tabular}
  }
  \vspace*{0.5cm}
\end{table*}

% -----------------------------------------------------------------------

\clearpage
% -----------------------------------------------------------------------
\twocolumn
\bibliography{applfast}
% -----------------------------------------------------------------------

\end{document}